\documentclass{doublecol}

\usepackage{amssymb}
\usepackage{amsbsy,natbib}
\usepackage{amsthm}
\usepackage{amsfonts,amsmath,bm}
\usepackage{epsfig}

\RequirePackage{graphics,epsf}

\usepackage{graphicx}
\usepackage[hang]{caption2}
\usepackage[normalsize,it,hang]{subfigure}
\usepackage{rotating}

\newcommand{\nom}{\ensuremath{\rm nom}}
\newcommand{\pure}{\ensuremath{\rm pure}}

\newtheorem{corollary}{Corollary}[section]
\newtheorem{proposition}{Proposition}[section]

\newtheorem{remark}{Remark}[section]

\newenvironment{notation}{\noindent{\bfseries Notation}}{\mbox{\ \\[1ex]}}

\newcommand{\Vect}[1]{ \boldsymbol{#1} }

\newcommand{\ZZZr}[1]{{\bfseries [#1]}}

\newcommand{\ZZZe}[1]{\marginpar{\vspace{-4mm}\small\raggedright
*}}
\renewcommand{\ZZZr}[1]{}

\renewcommand{\ZZZe}[1]{}

\begin{document}

\LRH{Non-linear estimation is easy}

\RRH{Michel Fliess, C\'edric Join and Hebertt Sira-Ram\'{\i}rez}

\VOL{1}

\ISSUE{1/2/3}

\PUBYEAR{2004}

\setcounter{page}{1}

\LRH{Non-linear estimation is easy}

\RRH{Michel Fliess, C\'{e}dric Join and Hebertt Sira-Ram\'{\i}rez}

\VOL{I}

\ISSUE{1/2/3}

\PUBYEAR{2004}

\BottomCatch

\title{Non-linear estimation is easy}

\authorA{Michel Fliess}

\affA{Projet ALIEN, INRIA Futurs \&
\'Equipe MAX, LIX (CNRS, UMR 7161), \'Ecole polytechnique, 91128
Palaiseau, France.\\
 \emph{E-mail}: {\tt Michel.Fliess}@{\tt polytechnique.edu}}

\authorB{C\'edric Join}

\affB{Projet ALIEN, INRIA Futurs \& CRAN (CNRS, UMR 7039),
Universit\'e Henri Poincar\'e (Nancy I), BP 239, 54506
Vand\oe{}uvre-l\`{e}s-Nancy, France.\\ \emph{E-mail}: {\tt
Cedric.Join}@{\tt cran.uhp-nancy.fr} }

\authorC{Hebertt Sira-Ram\'{\i}rez}

\affC{CINVESTAV-IPN, Secci\'on de Mecatr\'onica, Departamento de
Ingenier\'{\i}a El\'ectrica, Avenida IPN, No. 2508, Col. San Pedro
Zacatenco, AP 14740, 07300 M\'exico D.F., M\'exico.
\\ \emph{E-mail}: {\tt hsira}@{\tt cinvestav.mx}}

\begin{abstract}
Non-linear state estimation and some related topics, like parametric
estimation, fault diagnosis, and perturbation attenuation, are
tackled here via a new methodology in numerical differentiation. The
corresponding basic system theoretic definitions and properties are
presented within the framework of differential algebra, which
permits to handle system variables and their derivatives of any
order. Several academic examples and their computer simulations,
with on-line estimations, are illustrating our viewpoint.

\end{abstract}

\KEY{Non-linear systems, observability, parametric identifiability,
closed-loop state estimation, closed-loop parametric identification,
closed-loop fault diagnosis, closed-loop fault tolerant control,
closed-loop perturbation attenuation, numerical differentiation,
differential algebra.}


\BIO{M. Fliess is a Research Director at the {\em Centre National de
la Recherche Scientifique} and works at the {\em \'Ecole
Polytechnique} (Palaiseau, France). He is the head of the INRIA
project called ALIEN, which is devoted to the study and the
development of new techniques in identification and estimation. In
1991 he invented with J. L\'{e}vine, P. Martin, and P. Rouchon, the
notion of {\em differentially flat} systems which is playing a major
r\^{o}le in control applications.
\\ C. Join received his Ph.D. degree from the
University of Nancy, France, in 2002. He is now an Associate
Professor at the University of Nancy and is a member of the INRIA
project ALIEN. He is interested in the development of estimation
technics for linear and non-linear systems with a peculiar emphasis
in fault diagnosis and accommodation. His research involves also
signal and image processing. \\
H. Sira-Ram\'{\i}rez  obtained the Electrical Engineer's degree from the
{\em Universidad de Los Andes} in M\'{e}rida (Venezuela) in 1970. He
later obtained the MSc in EE and the Electrical Engineer degree, in
1974, and the PhD degree, also in EE, in 1977, all from the
Massachusetts Institute of Technology (Cambridge, USA). Dr.
Sira-Ram\'{\i}rez  worked for 28 years at the Universidad de Los Andes
where he held the positions of: Head of the Control Systems
Department, Head of the Graduate Studies in Control Engineering and
Vicepresident of the University. Currently,  he is a Titular
Researcher in the {\em Centro de Investigaci\'{o}n y Estudios Avanzados
del Instituto Polit\'{e}cnico Nacional} (CINVESTAV-IPN) in M\'{e}xico City
(M\'{e}xico). Dr Sira-Ram\'{\i}rez is a Senior Member of the Institute of
Electrical and Electronics Engineers (IEEE), a Distinguished
Lecturer from the same Institute and a Member of the IEEE
International Committee. He is also a member of the Society for
Industrial and Applied Mathematics (SIAM), of the International
Federation of Automatic Control (IFAC) and of the American
Mathematical Society (AMS). He is a coauthor of the books, {\em
Passivity Based Control of Euler-Lagrange Systems} published by
Springer-Verlag, in 1998, {\em Algebraic Methods in Flatness, Signal
Processing and State Estimation}, Lagares 2003, {\em Differentially
Flat Systems}, Marcel Dekker, 2004, {\em Control de Sistemas No
Lineales} Pearson-Prentice Hall 2006, and of {\em Control Design
Techniques in Power Electronics Devices}, Springer, 2006. Dr.
Sira-Ram\'{\i}rez is interested in the theoretical and practical aspects
of feedback regulation of nonlinear dynamic systems with special
emphasis in Variable Structure feedback control techniques and its
applications in Power Electronics.}

\maketitle

\section{Introduction}

\subsection{General overview}\label{go}

Since fifteen years non-linear flatness-based control
(\cite{flmr,flmr2}) has been quite effective in many concrete and
industrial applications (see also
\cite{lamna2,Rudolph03,sira03}). On the other hand, most of
the problems pertaining to non-linear state estimation, and to
related topics, like
\begin{itemize}

\item parametric estimation,

\item fault diagnosis and fault tolerant control,

\item perturbation attenuation,

\end{itemize}
remain largely open in spite of a huge literature\footnote{See,
e.g., the surveys and encyclopedia edited by
\cite{astrom,lamna1,lamna2,levine,menini,nijm,zinober}, and the
references therein.}. This paper aims at providing simple and
effective design methods for such questions. This is made possible
by the following facts: \\

\noindent According to the definition given by \cite{df1,df2}, a
non-linear input-output system is {\em observable} if, and only if,
any system variable, a state variable for instance, is a {\em
differential function} of the control and output variables, i.e., a
function of those variables and their derivatives up to some finite
order. This definition is easily generalized to parametric
identifiability and fault isolability. We will say more generally
that an unknown quantity may be determined if, and only if, it is
expressible as a differential function of the control and output
variables. \\

\noindent It follows from this conceptually simple and natural
viewpoint that non-linear estimation boils down to numerical
differentiation, i.e., to the derivatives estimations of noisy time
signals\footnote{The origin of flatness-based control may also be
traced back to a fresh look at controllability (\cite{smf}).}. This
classic ill-posed mathematical problem has been already attacked by
numerous means\footnote{For some recent references in the control
literature, see, e.g.,
\cite{braci,bus1,chitour,dabroom,diop-fromion,diop-grizzle,diop-grizzle-bis,duncan,ibrir0,ibrir,ibrir-diop,kelly,levant0,levant,su}.
The literature on numerical differentiation might be even larger in
signal processing and in other fields of engineering and applied
mathematics.}. We follow here another thread, which started in
\cite{nolcos} and \cite{compression,gretsi}: derivatives estimates
are obtained via integrations. This is the explanation of the quite
provocative title of this paper\footnote{There are of course
situations, for instance with a very strong corrupting noise, where
the present state of our techniques may be insufficient. See also
Remark \ref{chias}.} where non-linear asymptotic estimators are
replaced by differentiators, which are easy to
implement\footnote{Other authors like \cite{slotine} had already
noticed that ``good" numerical differentiators would greatly
simplify control synthesis.}.

\begin{remark}\label{remlin}
This approach to non-linear estimation should be regarded as an
extension of techniques for linear closed-loop parametric estimation
(\cite{fliess03,garnier}). Those techniques gave as a byproduct
linear closed-loop fault diagnosis (\cite{diag-lin}),
and linear state reconstructors (\cite{cras04}), which offer a
promising alternative to linear asymptotic observers and to Kalman's
filtering.
\end{remark}

\subsection{Numerical differentiation: a short summary of our
approach}\label{ins} .

Let us start with the first degree polynomial time function $p_1 (t)
= a_0 + a_1 t$, $t \geq 0$, $a_0, a_1 \in \mathbb{R}$. Rewrite
thanks to classic operational calculus (see, e.g.,
\cite{yosida}) $p_1$ as $P_1 = \frac{a_0}{s} +
\frac{a_1}{s^2}$. Multiply both sides by $s^2$:
\begin{equation}\label{1}
s^2 P_1 = a_0 s + a_1
\end{equation}
Take the derivative of both sides with respect to $s$, which
corresponds in the time domain to the multiplication by $- t$:
\begin{equation}\label{2}
s^2 \frac{d P_1}{ds} + 2s P_1 = a_0
\end{equation}
The coefficients $a_0, a_1$ are obtained via the triangular system
of equations (\ref{1})-(\ref{2}). We get rid of the time
derivatives, i.e., of $s P_1$, $s^2 P_1$, and $s^2 \frac{d
P_1}{ds}$, by multiplying both sides of Equations
(\ref{1})-(\ref{2}) by $s^{ - n}$, $n \geq 2$. The corresponding
iterated time integrals are low pass filters which attenuate the
corrupting noises, which are viewed as highly fluctuating phenomena
(cf. \cite{bruit}). A quite short time window is sufficient for
obtaining accurate values of $a_0$, $a_1$.

The extension to polynomial functions of higher degree is
straightforward. For derivatives estimates up to some finite order
of a given smooth function $f: [0, + \infty) \to \mathbb{R}$, take a
suitable truncated Taylor expansion around a given time instant
$t_0$, and apply the previous computations. Resetting  and utilizing
sliding time windows permit to estimate derivatives of various
orders at any sampled time instant.

\begin{remark}
Note that our differentiators
are not of asymptotic nature,
and do not require any statistical knowledge of the corrupting
noises.
Those two fundamental features remain therefore valid for our
non-linear estimation\footnote{They are also valid for the linear
estimation questions listed in Remark \ref{remlin}.}. This is a
change of paradigms when compared to most of today's
approaches\footnote{See, e.g., \cite{schweppe,jaulin}, and the references therein, for
other non-statistical approaches.}.
\end{remark}

\subsection{Analysis and organization of our paper}
Our paper is organized as follows. Section \ref{da} deals with the
differential algebraic setting for nonlinear systems, which was
introduced in \cite{fliess89,fliess90}. When compared to those
expositions and to other ones like
\cite{flmr,delaleau,Rudolph03,sira03}, the novelty lies in the two
following points:
\begin{enumerate}
\item The definitions of observability and parametric identifiability
are borrowed from \cite{df1,df2}.
\item We provide simple and natural definitions related to
non-linear diagnosis such as {\em detectability}, {\em isolability},
{\em parity equations}, and {\em residuals}, which are
straightforward extensions of the module-theoretic approach in
\cite{diag-lin} for linear systems.
\end{enumerate}
The main reason if not the only one for utilizing differential
algebra is the absolute necessity of considering derivatives of
arbitrary order of the system variables. Note that this could have
been also achieved with the differential geometric language of
infinite order prolongations (see, e.g.,
\cite{diff,flmr2})\footnote{The choice between the algebraic and
geometric languages is a delicate matter. The formalism of
differential algebra is perhaps suppler and more elegant, whereas
infinite prolongations permit to take advantage of the integration
of partial differential equations. This last point plays a crucial
r\^{o}le in the theoretical study of flatness (see, e.g.,
\cite{chet,martin1,martin2,murray,pomet,sastry}, and the references
therein) but seems to be unimportant here. Differential algebra on
the other hand permitted to introduce {\em quasi-static} state
feedbacks (\cite{silva1,silva2}), which are quite helpful in
feedback synthesis (see also \cite{dru,rd}). The connection of
differential algebra with constructive and computer algebra might be
useful in control (see, e.g., \cite{diop00,diop0,glad}, and the
references therein).}.

Section \ref{ns} details Subsection \ref{ins} on numerical differentiation.

Illustrations are provided by several academic
examples\footnote{These examples happen to be flat, although our
estimation techniques are not at all restricted to such systems. We
could have examined as well uncontrolled systems and/or non-flat
systems. The control of non-flat systems, which is much more
delicate (see, e.g., \cite{flmr,sira03}, and the references
therein), is beyond the scope of this article.} and their numerical
simulations\footnote{Any interested reader may ask C. Join for the
corresponding computer programs ({\tt Cedric.Join}@{\tt
cran.uhp-nancy.fr}).} which we wrote in a such a style that they are
easy to grasp without understanding the algebraic subtleties of
Section \ref{da}:
\begin{enumerate}
\item Section \ref{observer} is adapting a paper by \cite{fan} on a
non-linear observer. We only need for closing the loop derivatives
of the output signal. We nevertheless present also a state
reconstructor of an important physical variable.
\item Closed-loop parametric identification is achieved in Section
\ref{paramident}.
\item Section \ref{fault} deals with closed-loop fault diagnosis and
fault tolerant control.
\item Perturbation attenuation is presented in Section
\ref{pertatt}, via linear and non-linear case-studies.
\end{enumerate}
We end with a brief conclusion. First drafts of various parts of
this paper were presented in \cite{nolcos,zeitz}.

\section{Differential algebra}\label{da}
{\em Commutative algebra}, which is mainly concerned with the study
of commutative rings and fields, provides the right tools for
understanding algebraic equations (see, e.g.,
\cite{hartshorne,eisenbud}). {\em Differential algebra}, which was
mainly founded by \cite{ritt} and \cite{kolchin}, extends to
differential equations concepts and results from commutative
algebra\footnote{Algebraic equations are differential equations of
order $0$.}.
\subsection{Basic definitions} A {\em
differential ring} $R$, or, more precisely, an {\em ordinary
differential ring}, (see, e.g., \cite{kolchin} and
\cite{Chambert-Loir}) will be here a commutative ring\footnote{See,
e.g., \cite{ati,Chambert-Loir} for basic notions in commutative
algebra.} which is equipped with a single {\em derivation}
$\frac{d}{dt}: R \rightarrow R$ such that, for any $ a, b \in R$,
\begin{itemize}
\item $\frac{d}{dt} (a + b) = \dot{a} + \dot{b}$, \item
$\frac{d}{dt} (a b) = \dot{a}b + a \dot{b}$.
\end{itemize}
where $\frac{da}{dt} = \dot{a}$, $\frac{d^\nu a}{dt^\nu} =
a^{(\nu)}$, $\nu \geq 0$. A {\em differential field}, or, more
precisely, an {\em ordinary differential field}, is a differential
ring which is a field. A {\em constant} of $R$ is an element $c \in
R$ such that $\dot{c} = 0$. A {\em (differential) ring} (resp. {\em
field}) {\em of constants} is a differential ring (resp. field)
which only contains constants. The set of all constant elements of
$R$ is a subring (resp. subfield), which is called the {\em subring}
(resp.{\em subfield}) {\em of constants}.

A {\em differential ring} (resp. {\em field}) {\em extension} is
given by two differential rings (resp. fields) $R_1$, $R_2$, such
that
$R_1 \subseteq R_2$,
and qthe derivation of $R_1$ is the restriction to $R_1$ of the
derivation of $R_2$.

\begin{notation} Let $S$ be a subset of $R_2$. Write $R_1 \{ S \}$ (resp. $R_1
\langle S \rangle$) the differential subring (resp. subfield) of
$R_2$ generated by $R_1$ and $S$.
\end{notation}

\begin{notation} Let $k$ be a differential field and $X =
\{x_\iota | \iota \in I \}$ a set of {\em differential
indeterminates}, i.e., of indeterminates and their derivatives of
any order. Write $k \{X\}$ the differential ring of {\em
differential polynomials}, i.e., of polynomials belonging to $k [
x_{\iota}^{(\nu_\iota)} | \iota \in I; ~ \nu_\iota \geq 0 ]$. Any
differential polynomial is of the form $ \sum_{\mbox{\tiny \rm
finite}} c \prod_{\mbox{\tiny \rm finite}}
(x_{\iota}^{(\mu_\iota)})^{\alpha_{\mu_\iota}}$, $c \in k$.
\end{notation}

\begin{notation} If $R_1$ and $R_2$ are differential
fields, the corresponding field extension is often written $R_2 /
R_1$.
\end{notation}

A {\em differential ideal} $\frak{I}$ of $R$ is an ideal which is
also a differential subring. It is said to be {\em prime} if, and
only if, $\frak{I}$ is prime in the usual sense.

\subsection{Field extensions} All fields are assumed to be of
characteristic zero. Assume also that the differential field
extension $K/k$ is {\em finitely generated}, i.e., there exists a
finite subset $S \subset K$ such that $K = k\langle S \rangle$. An
element $a$ of $K$ is said to be {\em differentially algebraic} over
$k$ if, and only if, it satisfies an algebraic differential equation
with coefficients in $k$: there exists a non-zero polynomial $P$
over $k$, in several indeterminates, such that $P(a, \dot{a}, \dots,
a^{(\nu)}) = 0$. It is said to be {\em differentially
transcendental} over $k$ if, and only if, it is not differentially
algebraic. The extension $K/k$ is said to be {\em differentially
algebraic} if, and only if, any element of $K$ is differentially
algebraic over $k$. An extension which is not differentially
algebraic is said to be {\em differentially transcendental}.

The following result is playing an important r\^ole:
\begin{proposition}\label{dtr}
The extension $K/k$ is differentially algebraic if, and only if, its
transcendence degree is finite.
\end{proposition}

A set $\{ \xi_\iota \mid \iota \in I \}$ of elements in $K$ is said
to be {\em differentially algebraically independent} over $k$ if,
and only if, the set $\{ \xi^{(\nu)}_\iota \mid \iota \in I, \nu
\geq 0 \}$ of derivatives of any order is algebraically independent
over $k$. If a set is not differentially algebraically independent
over $k$, it is {\em differentially algebraically dependent over
$k$}. An independent set which is maximal with respect to inclusion
is called a {\em differential transcendence basis}. The
cardinalities, i.e., the numbers of elements, of two such bases are
equal. This cardinality is the {\em differential transcendence
degree} of the extension $K/k$; it is written $\mbox{\rm diff~tr~deg
} (K/k)$. Note that this degree is $0$ if, and only if, $K/k$ is
differentially algebraic.

\subsection{K\"ahler differentials}
{\em K\"ahler differentials} (see, e.g., \cite{hartshorne,eisenbud})
provide a kind of analogue of infinitesimal calculus in commutative
algebra. They have been extended to differential algebra by
\cite{johnson}. Consider again the extension $K/k$. Denote by
\begin{itemize}
\item $K[\frac{d}{dt}]$ the set of linear differential operators
$\sum_{\text{\rm finite}} a_\alpha \frac{d^\alpha}{dt^\alpha}$,
$a_\alpha \in K$, which is a left and right principal ideal ring
(see, e.g., \cite{McC});

\item $\Omega_{K/k}$ the left $K[\frac{d}{dt}]$-module of K\"ahler
differentials of the extension $K/k$;

\item $d_{K/k} x \in \Omega_{K/k}$ the (K\"ahler) differential of
$x \in K$.

\end{itemize}
\begin{proposition}\label{lineariz}
The next two properties are equivalent:

\begin{enumerate}
\item The set $\{x_\iota \mid \iota \in I \} \subset K$ is
differentially algebraically dependent (resp. independent) over $k$.

\item The set $\{d_{K/k}x_\iota \mid \iota \in I \}$ is
$K[\frac{d}{dt}]$-linearly dependent (resp. independent).
\end{enumerate}
\end{proposition}
The next corollary is a direct consequence from Propositions
\ref{dtr} and \ref{lineariz}.
\begin{corollary} The module $\Omega_{K/k}$ satisfies the
following properties:
\begin{itemize}
\item The {\em rank}\footnote{See, e.g., \cite{McC}.} of $\Omega_{K/k}$
is equal to the
differential transcendence degree of $K/k$.

\item $\Omega_{K/k}$ is {\em torsion}\footnote{See, e.g.,
\cite{McC}.} if, and only if, $K/k$ is differentially algebraic.

\item $\dim_{K} (\Omega_{K/k}) = \mbox{\rm tr}~\mbox{\rm deg}
(L/K)$. It is therefore finite if, and only if, $L/K$ is
differentially algebraic.

\item $\Omega_{K/k} = \{0\}$ if, and only if, $L/K$ is algebraic.
\end{itemize}

\end{corollary}

\subsection{Nonlinear systems}
\subsubsection{Generalities}\label{nlgen}
Let $k$ be a given differential ground field. A {\em (nonlinear)
(input-output) system} is a finitely generated differential
extension $K/k$. Set $K = k \langle S, \Vect{\textbf{W}},
\Vect{\pi} \rangle$ where

\begin{enumerate}
\item $S$ is a finite set of system variables, which contains the
sets $\Vect{u} = (u_1, \dots, u_m)$ and $\Vect{y} = (y_1, \dots,
y_p)$ of control and output variables,

\item $\Vect{\textbf{W}} = \{\textbf{w}_1, \dots, \textbf{w}_q\}$
denotes the {\em fault} variables,

\item $\Vect{\pi}= (\pi_1, \dots, \pi_r)$ denotes the {\em
perturbation}, or {\em disturbance}, variables.
\end{enumerate}
They satisfy the following properties:

\begin{itemize}
\item The control, fault and perturbation variables do not ``interact'',
i.e., the differential extensions $k\langle \Vect{u} \rangle / k$,
$k\langle \Vect{\textbf{W}} \rangle / k$ and $k\langle \Vect{\pi}
\rangle / k$  are {\em linearly disjoint}\footnote{See, e.g.,
\cite{eisenbud}.}.

\item The control (resp. fault) variables are
assumed to be {\em independent}, i.e., $\Vect{u}$ (resp.
$\Vect{\textbf{W}}$) is a differential transcendence basis of $k
\langle \Vect{u} \rangle /k$ (resp. $k \langle \Vect{\textbf{W}}
\rangle /k$).

\item The extension $K / k \langle \Vect{{u}}, \Vect{\textbf{W}},
\Vect{\pi} \rangle$ is differentially algebraic.

\item Assume that the differential ideal $( \Vect{\pi} )
\subset k\{ S, \Vect{\pi}, \Vect{\textbf{W}} \}$ generated by
$\Vect{\pi}$ is prime\footnote{Any reader with a good algebraic
background will notice a connection with the notion of {\em
differential specialization} (see, e.g., \cite{kolchin}).}. Write
$$k \{ S^{\text{nom}},
\Vect{\textbf{W}}^{\text{nom}} \} = k\{ S, \Vect{\pi},
\Vect{\textbf{W}} \} / ( \Vect{\pi} )$$ the quotient differential
ring, where the {\em nominal} system and fault variables
$S^{\text{nom}}$, $\Vect{\textbf{W}}^{\text{nom}}$ are the canonical
images of $S$, $\Vect{\textbf{W}}$. To those nominal variables
corresponds the {\em nominal system}\footnote{Let us explain those
algebraic manipulations in plain words. Ignoring the perturbation
variables in the original system yields the nominal system.}
$K^{\text{nom}} / k$, where $K^{\text{nom}} = k \langle
S^{\text{nom}}, \Vect{\textbf{W}}^{\text{nom}} \rangle$ is the
quotient field of $k \{ S^{\text{nom}},
\Vect{\textbf{W}}^{\text{nom}} \}$, which is an {\em integral
domain}, i.e., without zero divisors. The extension $K^{\text{nom}}
/ k \langle \Vect{u}^{\text{nom}}, \Vect{\textbf{W}}^{\text{nom}}
\rangle$ is differentially algebraic.

\item Assume as above that the differential ideal
$( \Vect{\textbf{W}}^{\text{nom}} )~\subset k\{ S^{\text{nom}} ,
\Vect{\textbf{W}}^{\text{nom}} \}$ generated by
$\Vect{\textbf{W}}^{\text{nom}}$ is prime. Write
$$k \{ S^{\text{pure}}\} = k\{ S^{\text{nom}} ,
\Vect{\textbf{W}}^{\text{nom}} \} / ( \Vect{\textbf{W}}^{\text{nom}}
)$$ where the {\em pure} system variables $S^{\text{pure}}$ are the
canonical images of $S^{\text{nom}}$. To those pure variables
corresponds the {\em pure system}\footnote{Ignoring as above the
fault variables in the nominal system yields the pure system.}
$K^{\text{pure}} / k$, where $K^{\text{pure}} = k \langle
S^{\text{pure}} \rangle$ is the quotient field of $k \{
S^{\text{pure}} \}$. The extension $K^{\text{pure}} / k \langle
\Vect{u}^{\text{pure}} \rangle$ is differentially algebraic.
\end{itemize}

\begin{remark}\label{difftr}
We make moreover the following natural assumptions: $\mbox{\rm
diff~tr~deg } (k \langle \Vect{{u}}^{\pure} \rangle /k) = \mbox{\rm
diff~tr~deg } (k \langle \Vect{{u}}^{\nom} \rangle /k) = \mbox{\rm
diff~tr~deg } (k \langle \Vect{{u}} \rangle /k) = m$, $\mbox{\rm
diff~tr~deg } (k \langle \Vect{\textbf{W}}^{\nom} \rangle /k) =
\mbox{\rm diff~tr~deg } (k \langle \Vect{\textbf{W}} \rangle /k) =
q$
\end{remark}

\begin{remark}
Remember that differential algebra considers algebraic differential
equations, i.e., differential equations which only contain
polynomial functions of the variables and their derivatives up to
some finite order. This is of course not always the case in
practice. In the example of Section \ref{observer}, for instance,
appears the transcendental function $\sin \theta_l$. As already
noted in \cite{flmr}, we recover algebraic differential equations by
introducing $\tan \frac{\theta_l}{2}$.
\end{remark}

\subsubsection{State-variable representation}

We know, from proposition \ref{dtr}, that the transcendence degree
of the extension $K / k \langle \Vect{\textbf{u}},
\Vect{\textbf{W}}, \Vect{\pi} \rangle$ is finite, say $n$. Let
$\Vect{x} = (x_1, \dots, x_n)$ be a transcendence basis. Any
derivative $\dot{x}_i$, $i = 1, \dots, n$, and any output variable
$y_j$, $j = 1,\dots, p$, are algebraically dependent over $k \langle
\Vect{\textbf{u}}, \Vect{\textbf{W}}, \Vect{\pi} \rangle$ on
$\Vect{x}$:
\begin{equation}
\label{state}
\begin{array}{l}
A_i (\dot{x}_i, \Vect{x}) = 0 \quad i = 1, \dots, n   \\
B_j (y_j, \Vect{x}) = 0 \quad j = 1, \dots, p
\end{array}
\end{equation}
where $A_i \in k \langle \Vect{\textbf{u}}, \Vect{\textbf{W}},
\Vect{\pi} \rangle [\dot{x}_i, \Vect{x}]$, $B_j \in k \langle
\Vect{\textbf{u}}, \Vect{\textbf{W}}, \Vect{\pi} \rangle [y_j,
\Vect{x}]$, i.e., the coefficients of the polynomials $A_i$, $B_j$
depend on the control, fault and perturbation variables and on their
derivatives up to some finite order.

Eq. (\ref{state}) becomes for the nominal system
\begin{equation}
\label{statenom}
\begin{array}{l}
A^{\text{nom}}_i (\dot{x}^{\text{nom}}_i, \Vect{x}^{\text{nom}}) = 0
\quad i = 1, \dots, n_{\text{nom}} \leq n  \\
B^{\text{nom}}_j (y^{\text{nom}}_j, \Vect{x}^{\text{nom}}) = 0 \quad
j = 1, \dots, p
\end{array}
\end{equation}
where $A^{\text{nom}}_i \in k \langle
\Vect{\textbf{u}}^{\text{nom}}, \Vect{\textbf{W}}^{\text{nom}}
\rangle [\dot{x}^{\text{nom}}_i, \Vect{x}^{\text{nom}}]$,
$B^{\text{nom}}_j \in k \langle \Vect{\textbf{u}}^{\text{nom}},
\Vect{\textbf{W}}^{\text{nom}} \rangle [y_{j}^{\text{nom}},
\Vect{x}^{\text{nom}}]$, i.e., the coefficients of
$A^{\text{nom}}_i$ and $B^{\text{nom}}_j$ depend on the nominal
control and fault variables and their derivatives and no more on the
perturbation variables and their derivatives.

We get for the pure system
\begin{equation}
\label{statepure}
\begin{array}{l}
A^{\text{pure}}_i (\dot{x}^{\text{pure}}_i, \Vect{x}^{\text{pure}}) = 0
\quad i = 1, \dots, n_{\text{pure}} \leq n_{\text{nom}}  \\
B^{\text{pure}}_j (y^{\text{pure}}_j, \Vect{x}^{\text{pure}}) = 0
\quad j = 1, \dots, p
\end{array}
\end{equation}
where $A^{\text{pure}}_i \in k \langle
\Vect{\textbf{u}}^{\text{pure}} \rangle [\dot{x}^{\text{pure}}_i,
\Vect{x}^{\text{pure}}]$, $B^{\text{pure}}_j \in k \langle
\Vect{\textbf{u}}^{\text{pure}} \rangle [y^{\text{pure}}_j,
\Vect{x}^{\text{pure}}]$, i.e., the coefficients of
$A^{\text{pure}}_i$ and $B^{\text{pure}}_j$ depend only on the pure
control variables and their derivatives.
\begin{remark}
Two main differences, which are confirmed by concrete examples (see,
e.g., \cite{hasler,grue}), can be made with the usual state-variable
representation
$$\begin{array}{l} \dot{\Vect{x}} = F(\Vect{x}, \Vect{u}) \\ \Vect{y} = H( \Vect{x} )
\end{array}$$
\begin{enumerate}
\item The representations (\ref{state}), (\ref{statenom}),
(\ref{statepure}) are implicit.
\item The derivatives of the control variables in the equations of the dynamics
cannot be in general removed (see \cite{dr}).
\end{enumerate}

\end{remark}

\subsection{Variational system\protect\footnote{See \cite{flmr} for more details.}}

Call $\Omega_{K/k}$ (resp. $\Omega_{K^{\text{nom}} /k}$,
$\Omega_{K^{\text{pure}} /k}$) the {\em variational}, or {\em
linearized}, {\em system} (resp. {\em nominal system}, {\em pure
system}) of system $K/k$. Proposition \ref{lineariz} yields for pure
systems
\begin{equation}
\label{transfer} A \left(\begin{array}{c}
d_{K^{\text{pure}} /k}y^{\text{pure}}_{1} \\
\vdots \\
d_{K^{\text{pure}} /k}y^{\text{pure}}_{p}
\end{array}
\right) = B \left(\begin{array}{c}
d_{K^{\text{pure}} /k}u^{\text{pure}}_{1} \\
\vdots \\
d_{K^{\text{pure}} /k}u^{\text{pure}}_{m}
\end{array}
\right)
\end{equation}
where\begin{itemize}

\item $A \in K [\frac{d}{dt}]^{p \times p}$ is of full rank,

\item $B \in K [\frac{d}{dt}]^{p \times m}$.

\end{itemize}
The {\em pure transfer matrix}\footnote{See \cite{lap} for more
details on transfer matrices of time-varying linear systems, and,
more generally, \cite{diag-lin}, \cite{bourles} for the
module-theoretic approach to linear systems.} is the matrix $A^{- 1}
B \in K(s)^{p \times m}$, where $K(s)$, $s = \frac{d}{dt}$, is the
skew quotient field\footnote{See, e.g., \cite{McC}.} of
$K[\frac{d}{dt}]$.

\subsection{Differential
flatness\protect\footnote{For more details see
\cite{flmr,Rudolph03,sira03}.}}

The system $K/k$ is said to be {\em (differentially)} flat if, and
only if, the pure system $K^{\text{pure}} /k$ is (differentially)
flat (\cite{flmr}): the algebraic closure $\bar{K}^{\text{pure}}$ of
$K^{\text{pure}}$ is equal to the algebraic closure of a purely
differentially transcendental extension of $k$. It means in other
words that there exists a finite subset $\Vect{z}^{\text{pure}} =
\{z_1^{\text{pure}}, \dots, z_m^{\text{pure}} \}$ of
$\bar{K}^{\text{pure}}$ such that
\begin{itemize}
\item $z_1^{\text{pure}}, \dots, z_m^{\text{pure}}$ are
differentially algebraically independent over $k$,

\item $z_1^{\text{pure}}, \dots, z_m^{\text{pure}}$ are
algebraic over $K^{\text{pure}}$,

\item any pure system variable is algebraic over
$k \langle z_1^{\text{pure}}, \dots, z_m^{\text{pure}} \rangle$.
\end{itemize}
$\Vect{z}^{\text{pure}}$ is a {\em (pure) flat}, or {\em
linearizing}, {\em output}. For a flat dynamics, it is known that
the number $m$ of its elements is equal to the number of independent
control variables.

\subsection{Observability and identifiability}
Take a system $K/k$ with control $\Vect{u}$ and output $\Vect{y}$.

\subsubsection{Observability}
According to \cite{df1,df2} (see also \cite{diop}), system $K/k$ is
said to be {\em observable} if, and only if, the extension
$K^{\text{pure}} / k \langle \Vect{u}^{\text{pure}},
\Vect{y}^{\text{pure}} \rangle$ is algebraic.

\begin{remark}
This new definition\footnote{See \cite{obsmfjr} for a definition via
infinite prolongations.} of observability is ``roughly" equivalent
(see \cite{df1,df2} for details\footnote{The differential algebraic
and the differential geometric languages are not equivalent. We
cannot therefore hope for a ``one-to-one bijection" between
definitions and results which are expressed in those two settings.})
to its usual differential geometric counterpart due to
\cite{hermann} (see also \cite{conte,gauthier,isidori,nij,sontag}).
\end{remark}

\subsubsection{Identifiable parameters\protect\footnote{Differential algebra
has already been employed for parametric identifiability and
identification but in a different context by several authors (see,
e.g., \cite{ljung,ollivier-these,sac}).}} Set $k = k_0 \langle
\mathbf{\Theta} \rangle$, where $k_0$ is a differential field and
$\mathbf{\Theta} =\{\theta_1, \dots, \theta_r \}$ a finite set of
{\em unknown parameters}, which might not be constant. According to
\cite{df1,df2}, a parameter $\theta_\iota$, $\iota = 1, \dots, r$,
is said to be {\em algebraically} (resp. {\em rationally}) {\em
identifiable} if, and only if, it is algebraic over (resp. belongs
to) $k_0 \langle \Vect{u}, \Vect{y} \rangle$:
\begin{itemize}
\item $\theta_\iota$ is rationally identifiable if, and only if,
it is equal to a differential rational function over $k_0$ of the
variables $\Vect{u}$, $\Vect{y}$, i.e., to a rational function of
$\Vect{u}$, $\Vect{y}$ and their derivatives up to some finite
order, with coefficients in $k_0$;

\item $\theta_\iota$ is algebraically identifiable if, and only
if, it satisfies an algebraic equation with coefficients in $k_0
\langle \Vect{u}, \Vect{y} \rangle$.
\end{itemize}

\subsubsection{{\em Determinable}
variables}\label{determin} More generally, a variable $\Upsilon \in
K$ is said to be {\em rationally} (resp. {\em algebraically}) {\em
determinable} if, and only if, $\Upsilon^{\text{pure}}$ belongs to
(resp. is algebraic over) $k \langle \Vect{u}^{\text{pure}},
\Vect{y}^{\text{pure}} \rangle$. A system variable $\chi$ is then
said to be {\em rationally} (resp. {\em algebraically}) {\em
observable} if, and only if, $\chi^{\text{pure}}$ belongs to (resp.
is algebraic over) $k \langle \Vect{u}^{\text{pure}},
\Vect{y}^{\text{pure}} \rangle$.

\begin{remark}\label{chias}
In the case of algebraic determinability, the corresponding
algebraic equation might possess several roots which are not easily
discriminated (see, e.g., \cite{chiasson} for a concrete example).
\end{remark}

\begin{remark}
See \cite{sedo} and \cite{ollivier} for efficient algorithms in
order to test observability and identifiability. Those algorithms
may certainly be extended to determinable variables and to various
questions related to fault diagnosis in Section \ref{faulty}.
\end{remark}

\subsection{Fundamental properties of fault variables\protect\footnote
{See, e.g., \cite{chen,blanke,gertler,vacht} for introductions to
this perhaps less well known subject. The definitions and properties
below are clear-cut extensions of their linear counterparts in
\cite{diag-lin}. Some of them might also be seen as a direct
consequence of Section \ref{determin}. Differential algebra has
already been employed but in a different context by several authors
(see, e.g., \cite{martinez,guerra,staros,zhang}).}}\label{faulty}

\subsubsection{Detectability} The fault variable
$\textbf{w}_\iota$, $\iota = 1, \dots, q$, is said to be {\em
detectable} if, and only if, the field extension $K^{\text{nom}} /
k \langle \Vect{u}^{\text{nom}},
\Vect{\textbf{W}}^{\text{nom}}_\iota \rangle$, where
$\Vect{\textbf{W}}^{\text{nom}}_\iota =
\Vect{\textbf{W}}^{\text{nom}} \backslash \{
\textbf{w}^{\text{nom}}_\iota\}$, is differentially
transcendental. It means that $\textbf{w}_\iota$ is indeed
``influencing" the output. When considering the variational
nominal system, formula (\ref{transfer}) yields
$$
\begin{array}{cl}\left( \begin{array}{c} d_{K^{\nom} / k}{y}_{1}^{\nom} \\ \vdots \\
d_{K^{\nom} / k}{y}_{p}^{\nom}
\end{array} \right) =& T_{\Vect{u}} \left( \begin{array}{c}
d_{K^{\nom} / k}{u}_{1}^{\nom} \\ \vdots \\
d_{K^{\nom} / k}{u}_{m}^{\nom}
\end{array} \right) \\&+ T_{\Vect{\textbf{W}}}
\left( \begin{array}{c} d_{K^{\nom} / k}{\textbf{w}}_{1}^{\nom} \\ \vdots \\
d_{K^{\nom} / k}{\textbf{w}}_{q}^{\nom}
\end{array} \right)\end{array}
$$
where $T_{\Vect{u}} \in K(s)^{p \times m}$, $T_{\Vect{\textbf{W}}}
\in K(s)^{p \times q}$. Call $T_{\Vect{\textbf{W}}}$ the {\em
fault transfer matrix}. The next result is clear:
\begin{proposition}\label{propo-detect}
The fault variable $\textbf{w}_\iota$ is detectable if, and only
if, the $\iota^{th}$ column of the fault transfer matrix
$T_{\Vect{\textbf{W}}}$ is non-zero.
\end{proposition}

\subsubsection{Isolability, parity equations and residuals}\label{pareq} A
subset $\Vect{\textbf{W}}^\prime = (\textbf{w}_{\iota_1}, \dots,
\textbf{w}_{\iota_{q^\prime}})$ of the set $\Vect{\textbf{W}}$ of
fault variables is said to be
\begin{itemize}

\item {\em Differentially algebraically isolable} if, and only if,
the extension $k \langle \Vect{u}^{\nom},\Vect{y}^{\nom},
\Vect{\textbf{W}}^{\prime \nom} \rangle / k \langle
\Vect{u}^{\nom},\Vect{y}^{\nom} \rangle$ is differentially
algebraic. It means that any component of $\Vect{\textbf{W}}^{\prime
\nom}$ satisfies a {\em parity differential equation}, i.e., an
algebraic differential equations where the coefficients belong to $k
\langle \Vect{u}^{\nom},\Vect{y}^{\nom} \rangle$.

\item {\em Algebraically isolable} if, and only if, the extension
$k \langle \Vect{u}^{\nom},\Vect{y}^{\nom},
\Vect{\textbf{W}}^{\prime \nom} \rangle / k \langle
\Vect{u}^{\nom},\Vect{y}^{\nom} \rangle$ is algebraic. It means that
the parity differential equation is of order $0$, i.e., it is an
algebraic equation with coefficients $k \langle
\Vect{u}^{\nom},\Vect{y}^{\nom} \rangle$.

\item {\em Rationally isolable} if, and only if,
$\Vect{\textbf{W}}^{\prime \nom}$ belongs to $k \langle
\Vect{u}^{\nom},\Vect{y}^{\nom} \rangle$. It means that the parity
equation is a linear algebraic equation, i.e., any component of
$\Vect{\textbf{W}}^{\prime \nom}$ may be expressed as a rational
function over $k$ in the variables $\Vect{u}^{\nom}$,
$\Vect{y}^{\nom}$ and their derivatives up to some finite order.
\end{itemize}
The next property is obvious:
\begin{proposition}
Rational isolability $\Rightarrow$ algebraic isolability
$\Rightarrow$ differentially algebraic isolability.
\end{proposition}
When we will say for short that fault variables are {\em
isolable}, it will mean that they are differentially algebraically
isolable.

\begin{proposition}
Assume that the fault variables belonging to
${\mbox\rm{\Vect{\textbf{W}}^\prime}}$ are isolable. Then
${\mbox{\rm card}} (\Vect{\textbf{W}}^{\prime}) \leq \mbox{\rm card}
(\Vect{y})$.
\end{proposition}

\begin{proof}
The differential transcendence degree of the extension $k \langle
\Vect{u}^{\nom},\Vect{y}^{\nom}, \Vect{\textbf{W}}^{\prime \nom}
\rangle / k$ (resp. $k \langle \Vect{u}^{\nom},\Vect{y}^{\nom}
\rangle / k$) is equal to $\mbox{\rm card} (\Vect{u}) + \mbox{\rm
card} (\Vect{W}^\prime)$ (resp. is less than or equal to $\mbox{\rm
card} (\Vect{u}) + \mbox{\rm card} (\Vect{y})$). The equality of
those two degrees implies our result thanks to the Remark
\ref{difftr}.
\end{proof}

%

\section{Derivatives of a noisy signal}\label{ns}

\subsection{Polynomial time signals}

Consider the real-valued polynomial function $x_{N} (t) = \sum_{\nu
= 0}^{N} x^{(\nu)}(0) \frac{t^\nu}{\nu !} \in \mathbb{R}[t]$, $t
\geq 0$, of degree $N$. Rewrite it in the well known notations of
operational calculus:
$$
X_N (s) = \sum_{\nu = 0}^{N} \frac{x^{(\nu)}(0)}{s^{\nu + 1}}
$$
We know utilize $\frac{d}{ds}$, which is sometimes called the {\em
algebraic derivative} (cf. \cite{miku1,miku2}). Multiply both sides
by $\frac{d^\alpha}{ds^\alpha} s^{N + 1}$, $\alpha = 0, 1, \dots,
N$. The quantities $x^{(\nu)}(0)$, $\nu = 0, 1, \dots, N$ are given
by the triangular system of linear equations\footnote{Following
\cite{fliess03,garnier}, those quantities are said to be {\em
linearly identifiable}.}:
\begin{equation}\label{triangu}
\frac{d^\alpha s^{N + 1} X_N}{ds^\alpha}  =
\frac{d^\alpha}{ds^\alpha} \left( \sum_{\nu = 0}^{N} x^{(\nu)}(0)
s^{N - \nu} \right)
\end{equation}
The time derivatives, i.e., $s^\mu \frac{d^\iota X_N}{ds^\iota}$,
$\mu = 1, \dots, N$, $0 \leq \iota \leq N$, are removed by
multiplying both sides of Eq. (\ref{triangu}) by $s^{- \bar{N}}$,
$\bar{N} > N$.
\begin{remark}
Remember (cf. \cite{miku1,miku2,yosida}) that $\frac{d}{ds}$
corresponds in the time domain to the multiplication by $- t$.
\end{remark}

\subsection{Analytic time signals}

Consider a real-valued analytic time function defined by the
convergent power series $x(t) = \sum_{\nu = 0}^{\infty}
x^{(\nu)}(0) \frac{t^\nu}{\nu !}$, where $0 \leq t < \rho$.
Introduce its truncated Taylor expansion
\begin{equation}\label{taylor}
x(t) = \sum_{\nu = 0}^{N} x^{(\nu)}(0) \frac{t^\nu}{\nu !} + O(t^{N
+ 1})
\end{equation}
Approximate $x(t)$ in the interval $(0, \varepsilon)$, $0 <
\varepsilon \leq \rho$, by its truncated Taylor expansion $x_{N} (t)
= \sum_{\nu = 0}^{N} x^{(\nu)}(0) \frac{t^\nu}{\nu !}$ of order $N$.
Introduce the operational analogue of $x(t)$, i.e., $X (s) =
\sum_{\nu \geq 0} \frac{x^{(\nu)}(0)}{s^{\nu + 1}}$, which is an
{\em operationally convergent} series in the sense of
\cite{miku1,miku2}. Denote by $[x^{(\nu)}(0)]_{e_N}(t)$, $0 \leq \nu
\leq N$, the numerical estimate of $x^{(\nu)}(0)$, which is obtained
by replacing $X_N(s)$ by $X (s)$ in Eq. (\ref{triangu}). The next
result, which is elementary from an analytic standpoint, provides a
mathematical justification for the computer implementations:
\begin{proposition}
For $0 < t < \varepsilon$,
\begin{equation}
\label{asympt} \lim_{t \downarrow 0} [x^{(\nu)}(0)]_{e_N}(t) =
\lim_{N \to +\infty} [x^{(\nu)}(0)]_{e_N}(t) = x^{(\nu)}(0)
\end{equation}
\end{proposition}
\begin{proof}
Following (\ref{taylor}) replace $x_N (t)$ by  $x(t) = x_N (t) +
O(t^{N + 1})$. The quantity $O(t^{N + 1})$ becomes negligible if $t
\downarrow 0$ or $N \to +\infty$.
\end{proof}

\begin{remark}
See \cite{mboup}) for fundamental theoretical developments. See also
\cite{nothen} for most fruitful comparisons and discussions.
\end{remark}

\subsection{Noisy signals}\label{noise}
Assume that our signals are corrupted by additive noises. Those
noises are viewed here as highly fluctuating, or oscillatory,
phenomena. They may be therefore attenuated by low-pass filters,
like iterated time integrals. Remember that those iterated time
integrals do occur in Eq. (\ref{triangu}) after multiplying both
sides by $s^{- \bar{N}}$, for $\bar{N} > 0$ large enough.

\begin{remark}\label{debruitage}
The estimated value of $x(0)$, which is obtained along those lines,
should be viewed as a denoising of the corresponding signal.
\end{remark}

\begin{remark}
See \cite{bruit} for a precise mathematical foundation, which is
based on {\em nonstandard analysis}. A highly fluctuating function
of zero mean is then defined by the following property: its integral
over a finite time interval is {\em infinitesimal}, i.e., ``very
small". Let us emphasize that this approach\footnote{This approach
applies as well to multiplicative noises (see \cite{bruit}). The
assumption on the noises being only additive is therefore
unnecessary.}, which has been confirmed by numerous computer
simulations and several laboratory experiments in control and in
signal processing\footnote{For numerical simulations in signal
processing, see \cite{compression,gretsi,gretsi-bis}. Some of them
are dealing with multiplicative noises.}, is independent of any
probabilistic setting. No knowledge of the statistical properties of
the noises is required.
\end{remark}

\section{Feedback and state reconstructor}\label{observer}
\subsection{System description}
Consider with \cite{fan} the mechanical system, depicted in Figure
\ref{sc-motor}. It consists of a DC-motor joined to an inverted
pendulum through a torsional spring:
\begin{figure}
\centering{\rotatebox{-0}{\includegraphics*[width=0.87\columnwidth]{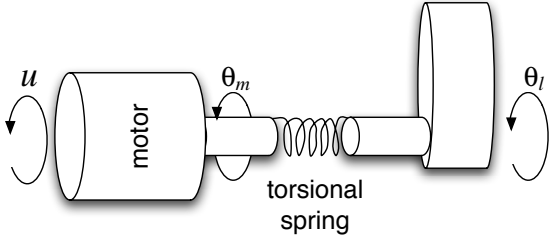}}}
 \caption{A single link flexible joint manipulator\label{sc-motor}}
 \end{figure}
\begin{equation}\label{nolcos}
\begin{array}{rl}
J_m\ddot{\theta}_m (t) &=\kappa\big(\theta_l (t) - \theta_m(t) \big)
-B\dot{\theta}_m (t) + K_\tau u (t) \\
J_l\ddot{\theta}_l (t) &=-\kappa\big(\theta_l(t) - \theta_m (t) \big)
-mgh\sin (\theta_l (t))\\
y(t) &=\theta_l (t)
\end{array}
\end{equation}
where
\begin{itemize}
\item $\theta_m$ and $\theta_l$ represent
respectively the angular deviation of the motor shaft and the
angular position of the inverted pendulum,
\item $J_m$, $J_l$, $h$, $m$, $\kappa$, $B$, $K_\tau$ and $g$ are
physical parameters which are assumed to be constant and known.
\end{itemize}
System (\ref{nolcos}), which is linearizable by static state
feedback, is flat; $y=\theta_l$ is a flat output.

\subsection{Control design}
Tracking of a given smooth reference trajectory
$y^\ast(t)=\theta^\ast_l(t)$ is achieved via the linearizing
feedback controller
\begin{equation}\label{feed1}
\begin{array}{rl}
u (t) &=\frac{1}{K_\tau}\Big(\frac{J_m}{\kappa}\big[J_lv(t) +\kappa\ddot{y}_e (t) \\
&+mgh(\ddot{y}_e(t) \cos (y_e (t)) -(\dot{y}_e(t))^2\sin (y_e(t)))\big]\\
&+J_l\ddot{y}_e(t) + mgh\sin (y_e(t))\\
&\frac{B}{\kappa}\big[J_ly^{(3)}_e (t) +\kappa\dot{y}_e (t) +
mgh\dot{y}_e (t) \cos (y_e(t) \big]\Big)
\end{array}
\end{equation}
where
\begin{equation}\label{feed2}
\begin{array}{rl}
v(t) &= y^{\ast (4)}(t) - \gamma_4(y_e^{(3)}(t) - y^{\ast (3)}(t))\\
&-\gamma_3(\ddot{y}_e (t) - \ddot{y}^\ast(t))-\gamma_2(\dot{y}_e (t) -
\dot{y}^\ast(t))\\
&-\gamma_1(y_e (t) -y^\ast(t))\\
\end{array}
\end{equation}
The subscript \textquotedblleft $e$\textquotedblright denotes the
estimated value. The design parameters $\gamma_1$, ..., $\gamma_4$
are chosen so that the resulting characteristic polynomial is
Hurwitz.
\begin{remark}
Feedback laws like (\ref{feed1})-(\ref{feed2}) depend, as usual in
flatness-based control (see, e.g., \cite{flmr,flmr2,sira03}), on the
derivatives of the flat output and not on the state variables.
\end{remark}

\subsection{A state reconstructor\protect\footnote{See \cite{chaos} and \cite{reger}
for other interesting examples of state reconstructors which are
applied to chaotically encrypted messages.}} We might nevertheless
be interested in obtaining an estimate $[\theta_m]_e (t)$ of the
unmeasured state $\theta_m (t)$:
\begin{equation}
[\theta_m]_e (t) =\frac{1}{\kappa}\Big(J_l\ddot{y}_e (t) + mgh\sin
(y_e (t))\Big) + y_e (t) \label{x-DC}
\end{equation}

\subsection{Numerical simulations}
\begin{figure}
{\rotatebox{-90}{\includegraphics*[width=0.81\columnwidth]{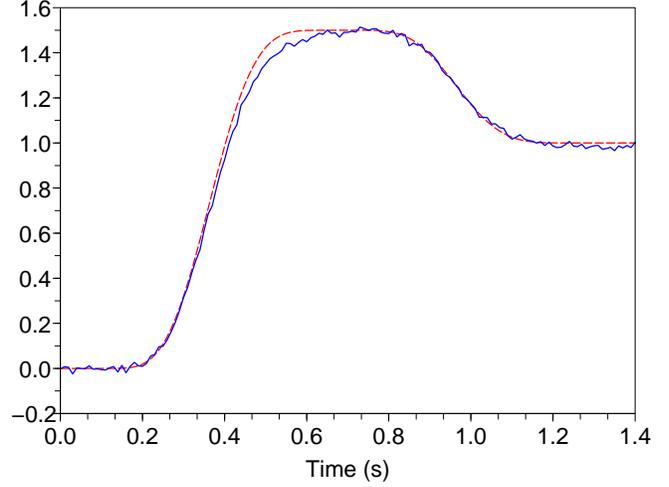}}}
\caption{Output (--) and reference trajectory (- -)} \label{join1}
\end{figure}

\begin{figure}
{\rotatebox{-90}{\includegraphics*[width=0.81\columnwidth]{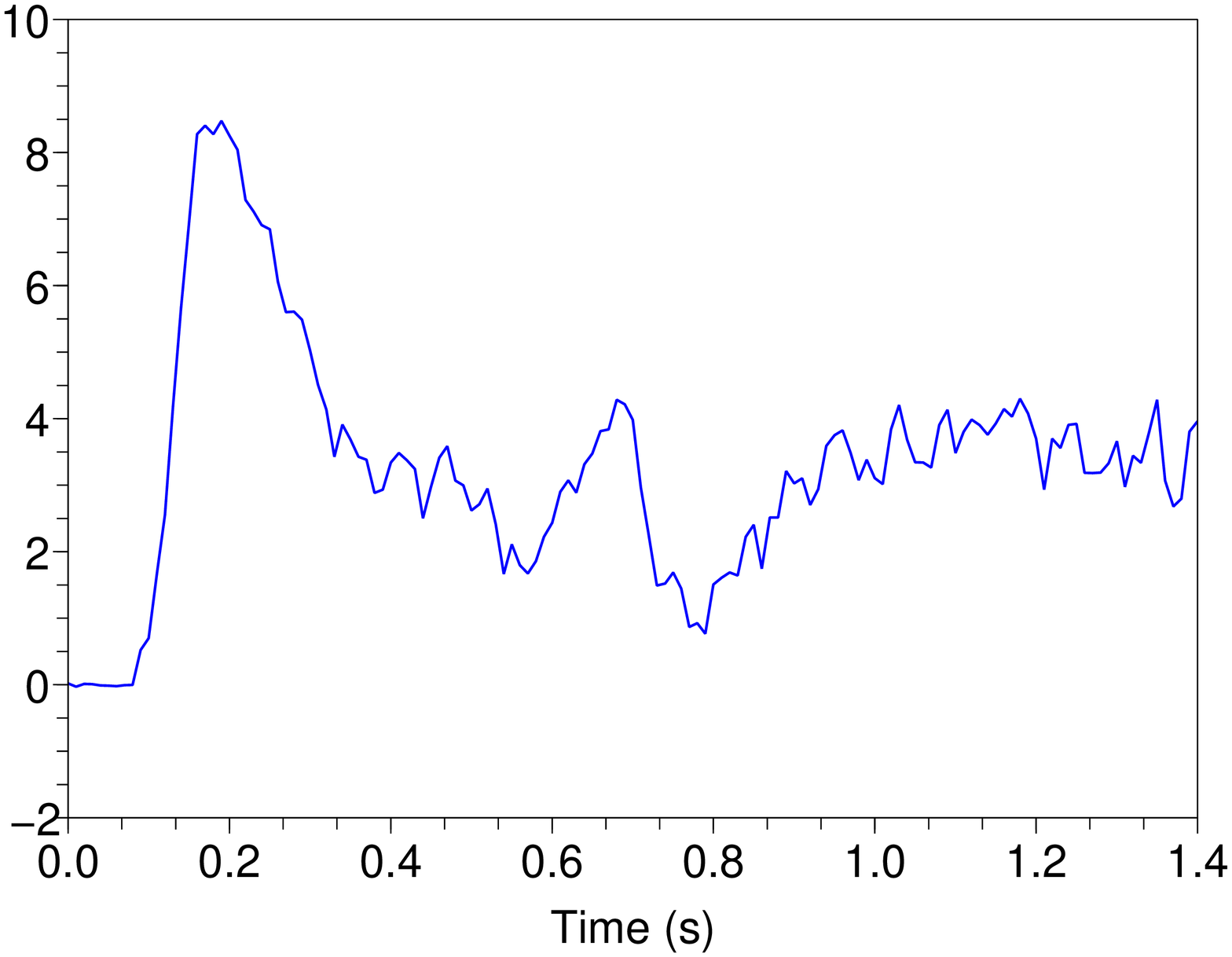}}}
\caption{Control} \label{join1bis}
\end{figure}

\begin{figure}
{\rotatebox{-90}{\includegraphics*[width=0.81\columnwidth]{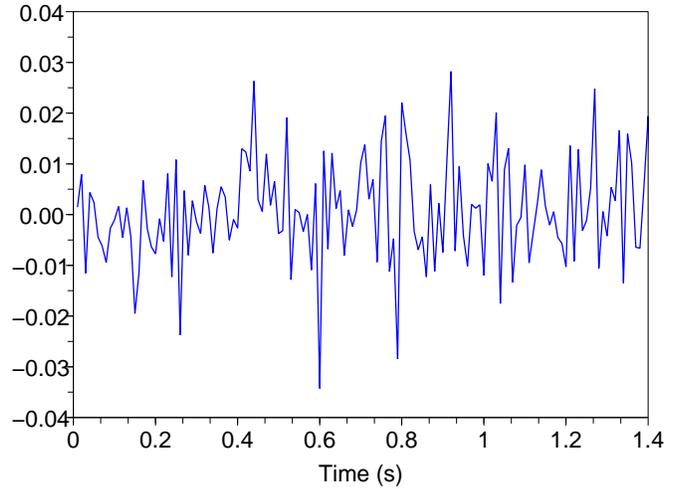}}}
\caption{Output noise} \label{join2}
\end{figure}

\begin{figure*}
{\rotatebox{-90}{\includegraphics*[width=1.25\columnwidth]{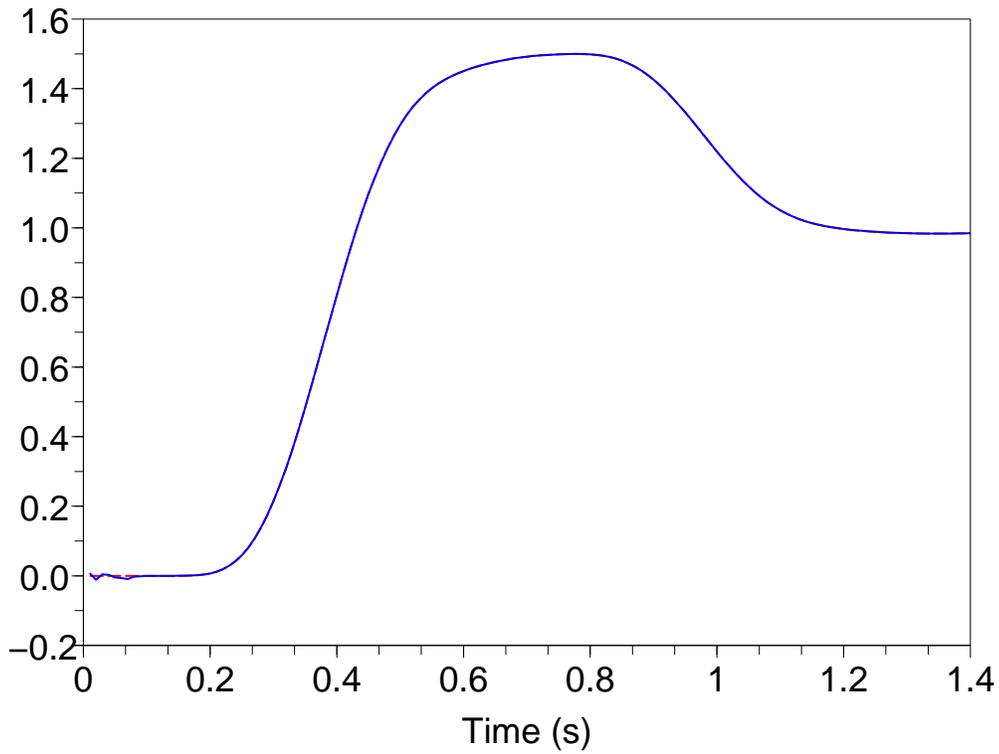}}}
\caption{$y$: (- -); on-line noise attenuation $y_e$ (--)}
\label{join3}
\end{figure*}

\begin{figure*}
{\rotatebox{-90}{\includegraphics*[width=1.25\columnwidth]{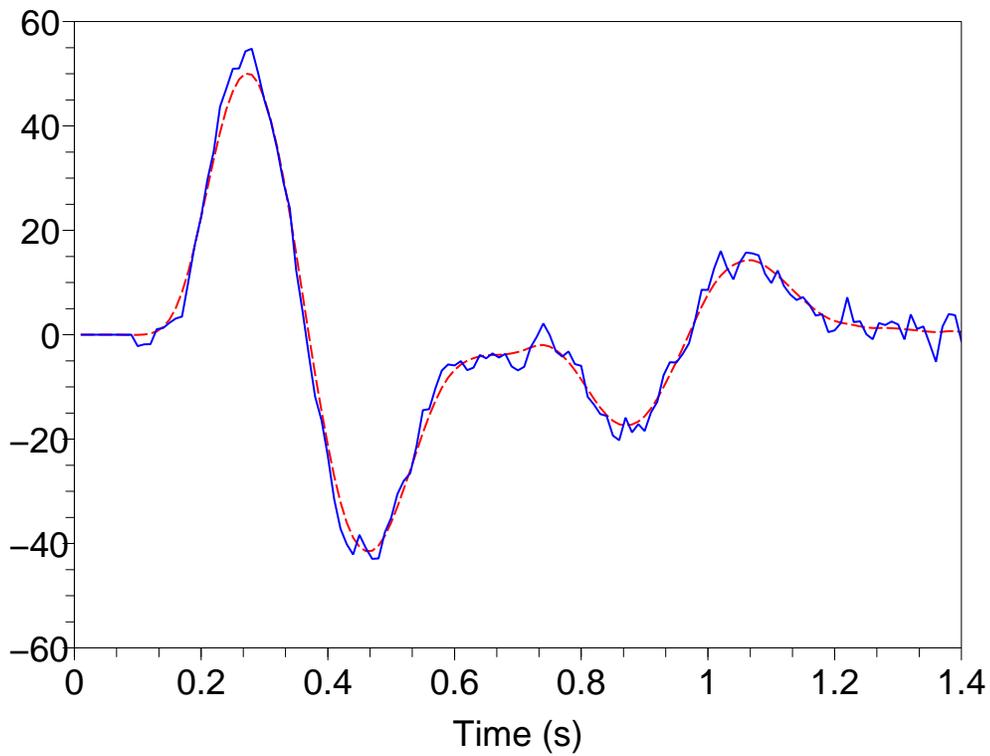}}}
\caption{$\ddot y$ (- -); on-line estimation $\ddot y_e$ (--)},
\label{join4}
\end{figure*}

\begin{figure*}
{\rotatebox{-90}{\includegraphics*[width=1.25\columnwidth]{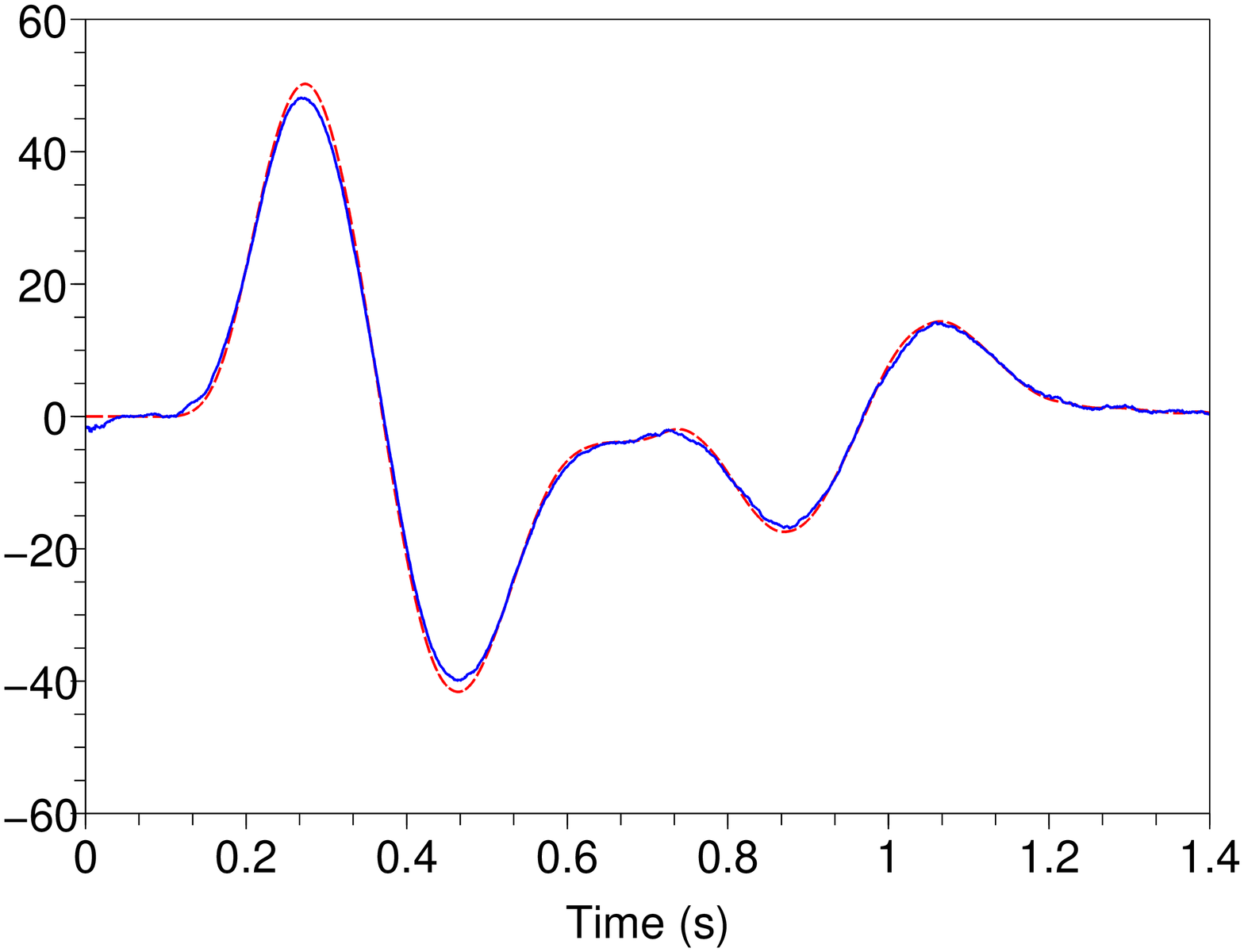}}}
\caption{$\ddot y$ (- -); off-line estimation $\ddot y_e$ (--) }
\label{join6}
\end{figure*}

\begin{figure*}
{\rotatebox{-90}{\includegraphics*[width=1.25\columnwidth]{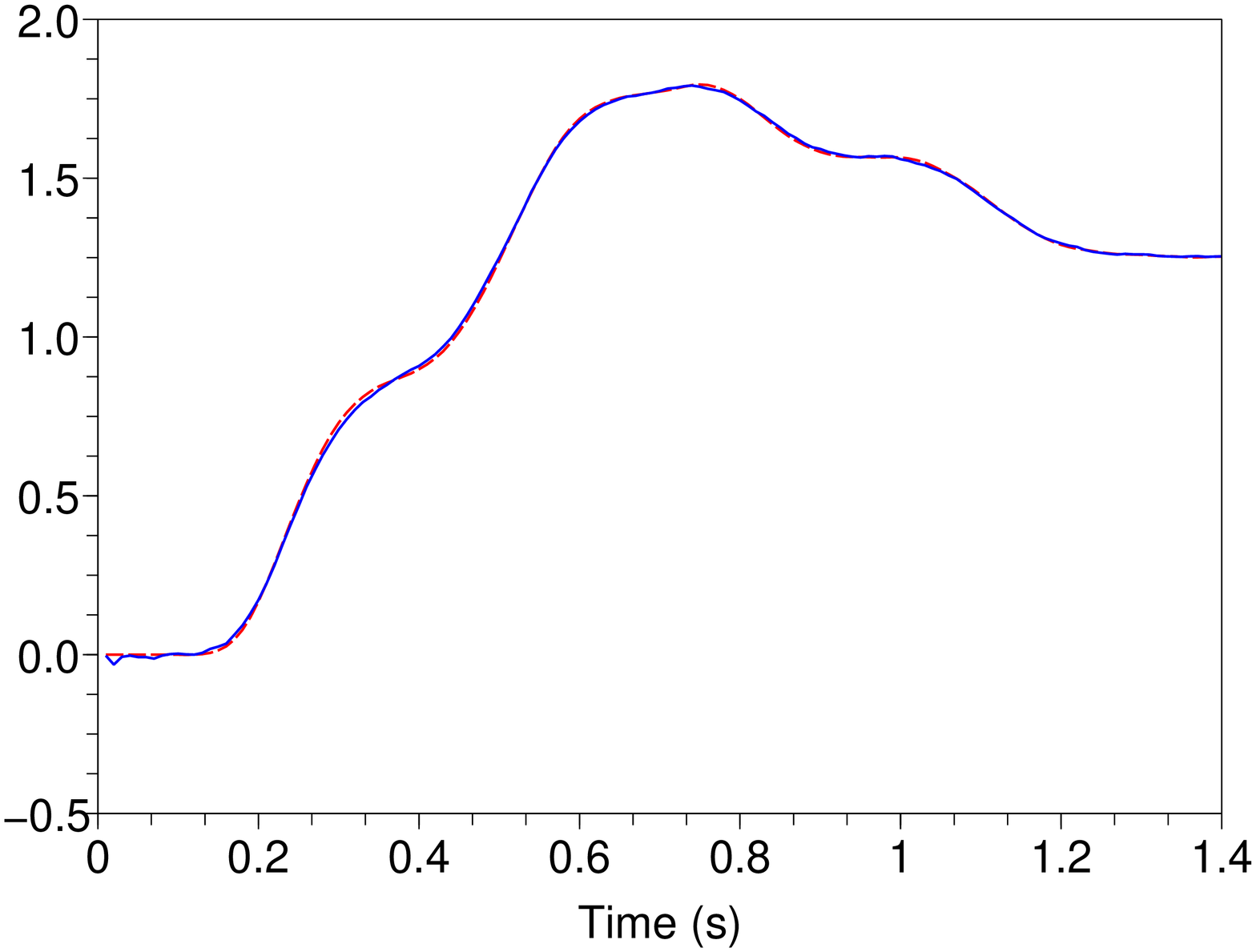}}}
\caption{$\theta_m$ (- -); off-line estimation
$\text{[}\theta_m\text{]}_e$ (--)} \label{join7}
\end{figure*}

The physical parameters have the same numerical values as in
\cite{fan}: $J_m=3.7\times10^{-3}\textit{ kgm$^2$}$,
$J_l=9.3\times10^{-3}\textit{ kgm$^2$}$, $h=1.5\times10^{-1}\textit{
m}$, $m=0.21\textit{ kg}$, $B=4.6\times10^{-2}\textit{ m}$,
$K_\tau=8\times10^{-2}\textit{ NmV$^{-1}$}$. The numerical
simulations are presented in Figures \ref{join1} - \ref{join7}.
Robustness has been tested with an additive white Gaussian noise
N(0; 0.01) on the output $y$. Note that the off-line estimations of
$\ddot y$ and $\theta_m$, where a ``small" delay is allowed, are
better than the on-line estimation of $\ddot y$.


\begin{figure*}\label{join5}
{\rotatebox{-90}{\includegraphics*[width=1.25\columnwidth]{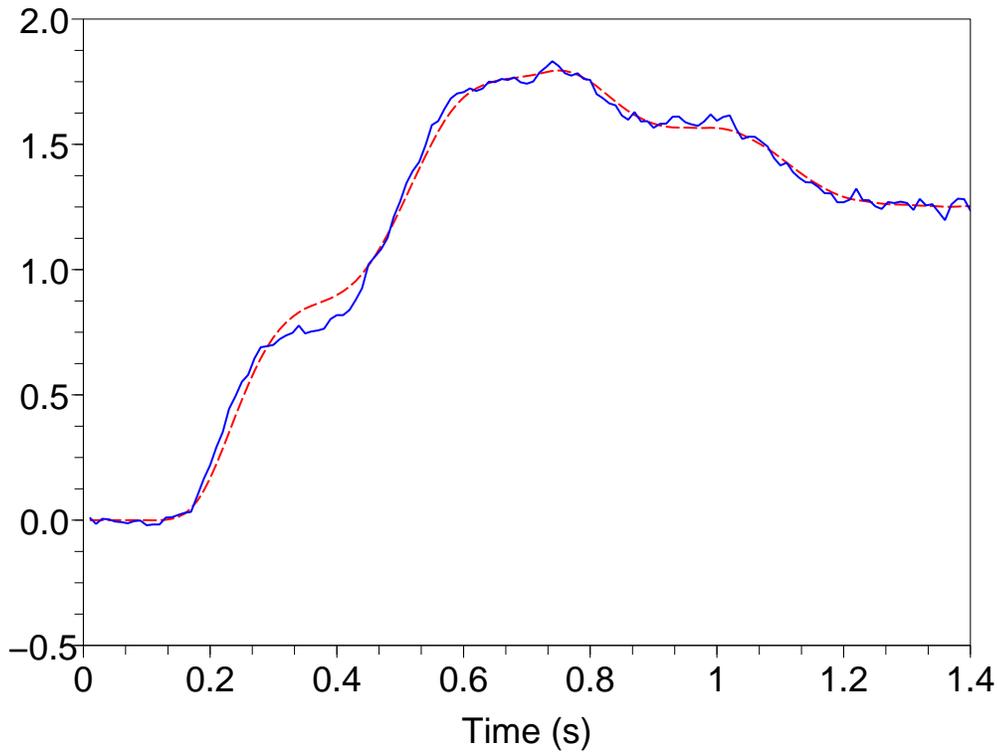}}}
\caption{$\theta_m$ (- -); on-line estimation
$\text{[}\theta_m\text{]}_e$ (--)} \label{join5}
\end{figure*}

\section{Parametric identification}\label{paramident}
\subsection{A rigid body}
\begin{figure}
\centering{\rotatebox{-0}{\includegraphics*[width=0.85\columnwidth]{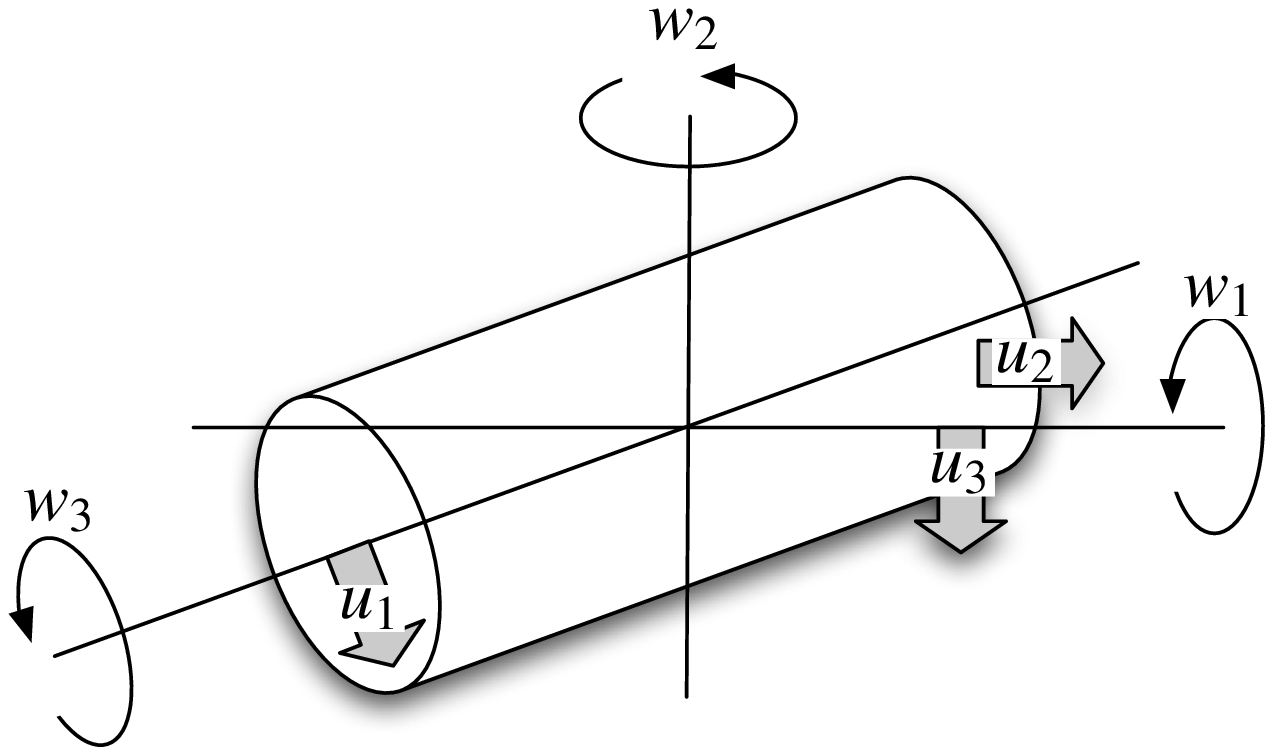}}}
 \caption{Rigid body\label{sc-euler}}
 \end{figure}

Consider the fully actuated rigid body, depicted in Figure
\ref{sc-euler}, which is given by the Euler equations
\begin{equation}\label{rigid}
\begin{array}{rl}
I_1\dot{w}_1 (t) = (I_2-I_3)w_2(t) w_3(t) + u_1(t) \\
I_2\dot{w}_2 (t) = (I_3-I_1)w_3 (t) w_1(t) +u_2 (t) \\
I_3\dot{w}_3 (t) = (I_1-I_2)w_1 (t) w_2 (t) + u_3 (t) \\
\end{array}
\end{equation}
where $w_1$, $w_2$, $w_3$ are the measured angular velocities,
$u_1$, $u_2$, $u_3$ the applied control input torques, $I_1$, $I_2$,
$I_3$ the constant moments of inertia, which are poorly known.
System (\ref{rigid}) is stabilized around the origin, for suitably
chosen design parameters $\lambda_{1\iota}$, $\lambda_{0\iota}$,
$\iota= 1,2,3$, by the feedback controller, which is an obvious
extension of the familiar proportional-integral (PI) regulators,
\begin{equation}\label{euler}
\begin{array}{rl}
u_1(t) =&-(I_2-I_3)w_2(t) w_3(t)\\&+I_1\Big( -\lambda_{11}w_1(t)
-\lambda_{01}\int_0^t w_1(\sigma)d\sigma \Big)\\
u_2(t) =&-(I_3-I_1)w_3(t) w_1(t) \\&+I_2\Big( -\lambda_{12}w_2(t)
-\lambda_{02}\int_0^tw_2(\sigma)d\sigma \Big)\\
u_3(t) =&-(I_1-I_2)w_1(t) w_2(t) \\&+I_3\Big( -\lambda_{13}w_3(t)
-\lambda_{03}\int_0^tw_3(\sigma)d\sigma \Big)\\
\end{array}
\end{equation}

\subsection{Identification of the moments of inertia}
Write Eq. (\ref{rigid}) in the following matrix form:
\begin{equation*}\label{inertia}
\begin{array}{l}
\begin{pmatrix}
\dot{w}_1 & - w_2 w_3 & w_2 w_2 \\
w_1 w_3 & \dot{w}_2 & - w_1 w_3 \\
- w_1 w_2  & w_1 w_2  & \dot{w}_3
\end{pmatrix} \times
\\
\begin{pmatrix}
I_1 \\
I_2 \\
I_3
\end{pmatrix}=
\begin{pmatrix}
u_1\\
u_2\\
u_3
\end{pmatrix}
\end{array}
\end{equation*}
It yields estimates
$[I_1]_e$, $[I_2]_e$, $[I_3]_e$ of $I_1$, $I_2$, $I_3$ when we
replace
$w_1$, $w_2$, $w_3$, $\dot{w}_1$, $\dot{w}_2$, $\dot{w}_3$ by their
estimates\footnote{See Remark \ref{debruitage}.}. The control law
(\ref{euler}) becomes
{\small \begin{equation}\label{euler-bis}
\begin{array}{rl} u_1(t)=&-([I_2]_e-[I_3]_e)[w_2]_e(t)
[w_3]_e(t)\\&+[I_1]_e\Big( -\lambda_{11}[w_1]_e(t)
-\lambda_{01}\int_0^t[w_1]_e(\sigma)d\sigma \Big)\\
u_2(t) =&-([I_3]_e-[I_1]_e)[w_3]_e(t) [w_1]_e(t)\\&+[I_2]_e\Big( -\lambda_{12}[w_2]_e(t)
-\lambda_{02}\int_0^t[w_2]_e(\sigma)d\sigma \Big)\\
u_3(t) =&-([I_1]_e-[I_2]_e)[w_1]_e(t) [w_2]_e(t) \\&+[I_3]_e\Big( -\lambda_{13}[w_3]_e(t)
-\lambda_{03}\int_0^t[w_3]_e(\sigma)d\sigma \Big)\\
\end{array}
\end{equation}}

\subsection{Numerical simulations}
\begin{figure*}
\centering{\rotatebox{-90}{\includegraphics*[width=1.25\columnwidth]{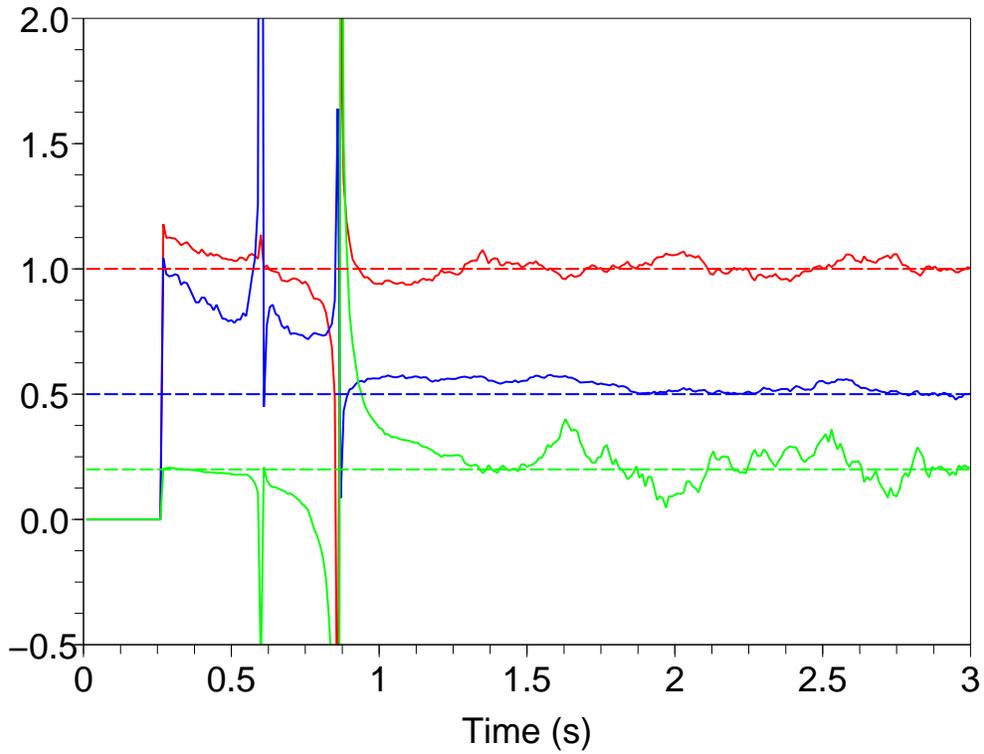}}}
\caption{Zoom on the parametric estimation (--) and real values (-
-)\label{E-est}}
\end{figure*}

\begin{figure*}
\centering{\rotatebox{-90}{\includegraphics*[width=1.25\columnwidth]{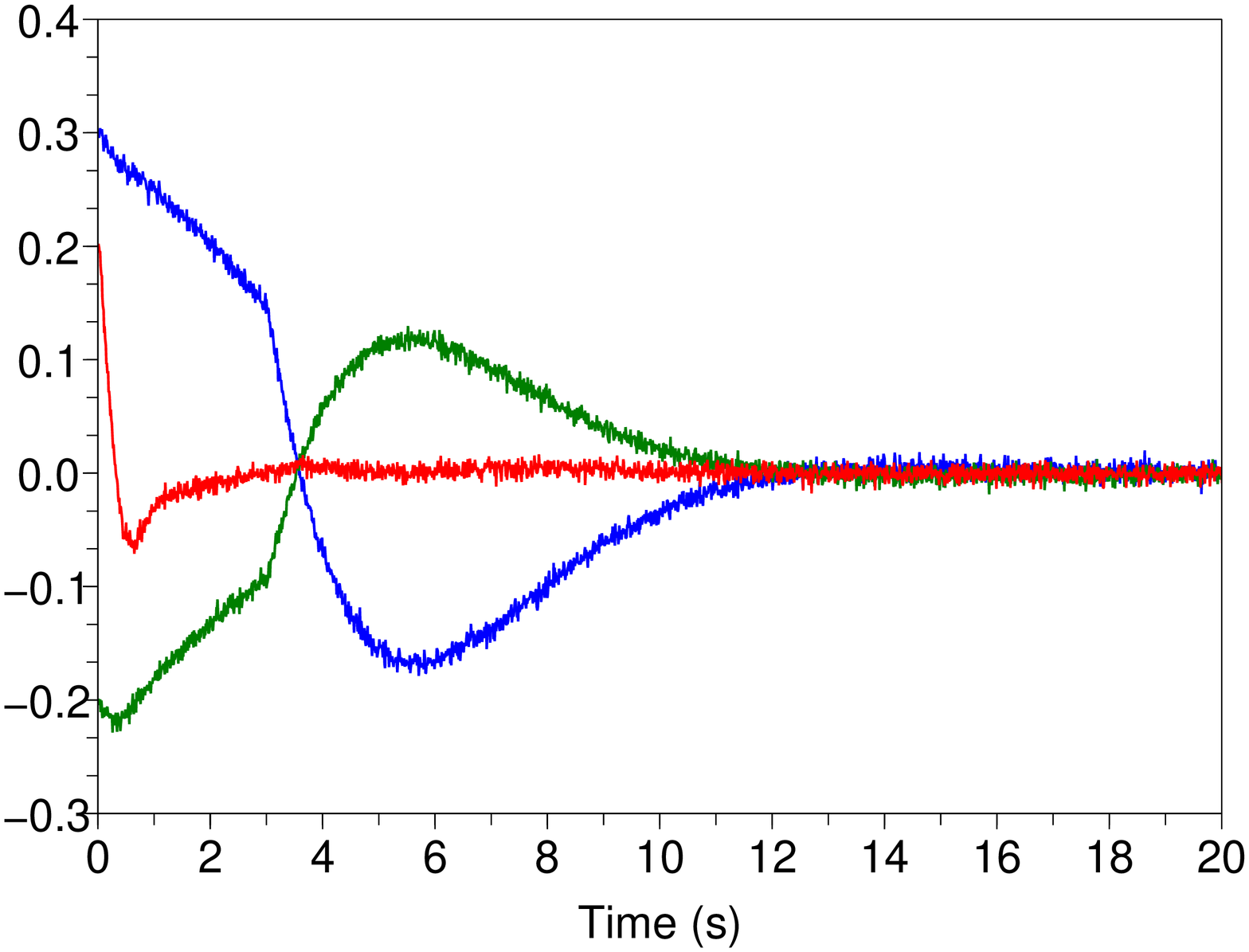}}}
\caption{Feedback stabilization with parametric
estimation\label{E-avec}}
\end{figure*}

\begin{figure}
\centering{\rotatebox{-90}{\includegraphics*[width=0.8\columnwidth]{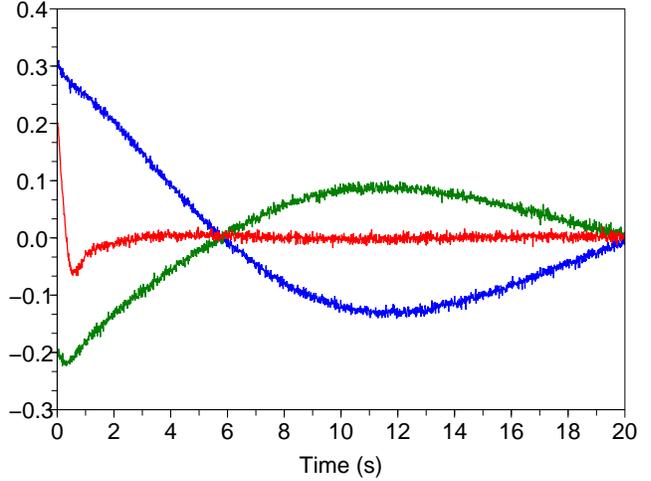}}}
\caption{Feedback stabilization without parametric
estimation\label{E-sans}}
\end{figure}

The output measurements are corrupted by an additive Gaussian white
noise $N(0; 0.005)$. Figure \ref{E-est} shows an excellent on-line
estimation of the three moments of inertia. Set for the design
parameters in the controllers (\ref{euler}) and (\ref{euler-bis})
$\lambda_{1\iota}=2\xi \varpi$, $\lambda_{0\iota}=\varpi^2$, $\iota
= 1, 2, 3$, where $\xi = 0.707$, $\varpi = 0.5$. The stabilization
with the above estimated values in Figure \ref{E-avec} is quite
better than in Figure \ref{E-sans} where the following false values
where utilized: $I_1 =0.2$, $I_2=0.1$ and $I_3 = 0.1$.

\section{Fault diagnosis and accommodation}\label{fault}
\subsection{A two tank system\protect\footnote{See \cite{mai} for another example.} }
\begin{figure}[h]
\begin{center}
\epsfig{height=3in, width=3in, file=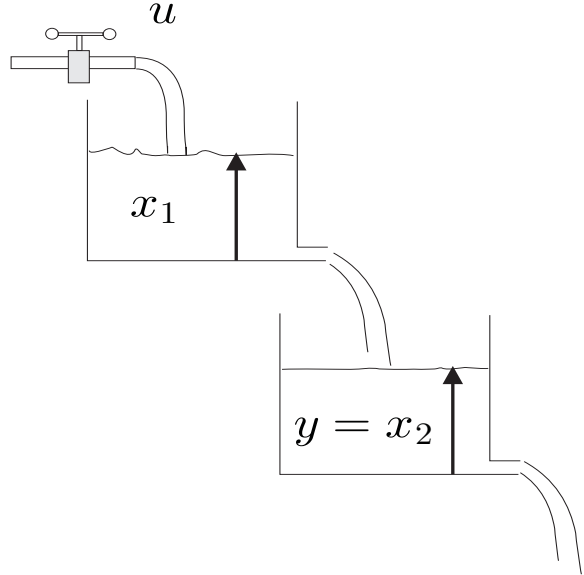}
\caption{A two tank system}\label{fig:7.12}
\end{center}
\end{figure}

Consider the cascade arrangement of two identical tank systems,
shown in Figure \ref{fig:7.12}, which is a popular example in fault
diagnosis (see, e.g., \cite{blanke}).

\noindent Its mathematical description is given by
\begin{eqnarray}
{\dot x}_1 (t) &=& -\frac{c}{A}\sqrt{x_1 (t)}+\frac{1}{A}u(t)
\left(1-{\bf w}(t) \right) \nonumber \\
&& +\varpi (t)  \label{tank:2} \\
{\dot x}_2 (t) &=& \frac{c}{A}\sqrt{x_1 (t) }-\frac{c}{A}\sqrt{x_2 (t)} \nonumber\\
y(t) &=& x_2 (t) \nonumber
\end{eqnarray}
where:
\begin{itemize}
\item The constant $c$ and the area $A$ of the tank's bottom are
known parameters.
\item The perturbation $\varpi (t)$ is constant but unknown,
\item The actuator failure ${\bf w}(t)$, $0 \leq {\bf w}(t) \leq 1$, is constant but
unknown. It starts at some unknown time $t_I > > 0$ which is not
``small".
\item Only the output $y=x_2$ is available for measurement.
\end{itemize}
The corresponding pure system, where we are ignoring the fault and
perturbation variables (cf. Section \ref{nlgen}),
\begin{equation}
\begin{array}{l}
\dot x^{\text{pure}}_1 = -\frac{c}{A}\sqrt{x^{\text{pure}}_1}+\frac{1}{A}u^{\text{pure}} \nonumber\\
\dot x^{\text{pure}}_2 = \frac{c}{A}\sqrt{x^{\text{pure}}_1} -
\frac{c}{A}\sqrt{x^{\text{pure}}_2}\nonumber\\
y^{\text{pure}} = x^{\text{pure}}_2 
\end{array}
\end{equation}
is flat. Its flat output is $y^{\text{pure}} = x^{\text{pure}}_2$.
The state variable $x^{\text{pure}}_1$ and control variable
$u^{\text{pure}}$ are given by
\begin{eqnarray}
x^{\text{pure}}_1 &=& \big(\frac{A}{c}{\dot y}^{\text{pure}} +
\sqrt{y^{\text{pure}}}\big)^2  \label{pur-plat1} \\
u^{\text{pure}} &=& 2A\big(\frac{A}{c}{\dot y}^{\text{pure}}
+\sqrt{y^{\text{pure}}}\big)\big(\frac{A}{c}{\ddot y}^{\text{pure}}
+ \frac{{\dot y}^{\text{pure}}}{2\sqrt{y^{\text{pure}}}}\big)
\nonumber \\
&& +c\big(\frac{A}{c}{\dot y}^{\text{pure}}
+\sqrt{y^{\text{pure}}}\big) \label{pur-plat2}
\end{eqnarray}

\subsection{Fault tolerant tracking controller}\label{controller}
It is desired that the output $y$ tracks a given smooth reference
trajectory $y^{\ast}(t)$. Rewrite Formulae
(\ref{pur-plat1})-(\ref{pur-plat2}) by taking into account the
perturbation variable $\varpi (t)$ and the actuator failure ${\bf
w}(t)$:
\begin{eqnarray}
x_1 (t) &=& \big(\frac{A}{c}{\dot y}(t) + \sqrt{y(t)}\big)^2 \label{x1} \\
u (t) &=& \frac{1}{\big(1-{\bf w}(t)\big)}\Big(-A\varpi\nonumber\\
&&+2A\big(\frac{A}{c}{\dot y} (t) +\sqrt{y (t)}\big)\big. \times
\nonumber \\ && (\frac{A}{c}{\ddot y} (t) + \frac{{\dot y}
(t)}{2\sqrt{y (t)}}\big) \nonumber \\ && +c\big(\frac{A}{c}{\dot
y}(t) +\sqrt{y (t)}\big) \Big)\nonumber
\end{eqnarray}
With reliable on-line estimates $\hat{\bf w}(t)$ and $\hat{\varpi}
(t)$ of the failure signal ${\bf w}(t)$ and of the perturbation
$\varpi (t)$, we design a failure accommodating linearizing feedback
controller. It incorporates a classical robustifying integral
action:
\begin{eqnarray}
u(t) &=& \frac{1}{\big(1-{\hat{\bf w}}(t)\big)}\Big(-A{\hat{\varpi}} (t) \nonumber\\
&&+2A\big(\frac{A}{c}{\dot y}_e(t) +\sqrt{y_e
(t)}\big)\big(\frac{A}{c}v (t) +\frac{{\dot y}_e (t)}{2\sqrt{y_e
(t)}}\big) \nonumber\\
&& +c\big(\frac{A}{c}{\dot y}_e (t) +\sqrt{y_e (t)}\big) \Big)  \nonumber\\
v (t) &=&  {\ddot y^{\ast}} (t) - \mathcal{G} \star (y_e (t) -
y^{\ast}(t))\nonumber
\end{eqnarray}
This is a {\em generalized proportional integral (GPI)} controller
(cf. \cite{gpi}) where
\begin{itemize}
\item $\star$ denotes the convolution product,
\item the transfer function of $\mathcal{G}$ is
$$\frac{\lambda_2s^2 + \lambda_1s + \lambda_0}{s(s+\lambda_3)}$$
where $\lambda_0, \lambda_1, \lambda_2, \lambda_3 \in \mathbb{R}$,
\item $y_e (t)$ is the on-line denoised estimate of $y(t)$ (cf. Remark \ref{debruitage}),
\item ${\dot y}_e (t)$ is the on-line estimated
value of $\dot{y} (t)$.
\end{itemize}

\subsection{Perturbation and fault estimation}
The estimation of the constant perturbation $\varpi$ is readily
accomplished from Eq. (\ref{tank:2}) before the occurrence of the
failure ${\bf w}$, which starts at time $t_I >> 0$:
$$
\dot x_1 (t) = -\frac{c}{A}\sqrt{x_1 (t)}+ \frac{1}{A}u (t) + \varpi
\quad \mbox{\rm if}~ ~ 0 < t < t_I
$$
Multiplying both sides by $t$ and integrating by parts
yields\footnote{We are adapting here linear techniques stemming from
\cite{fliess03,garnier}.}
\[
{\hat \varpi}=\left\{\begin{array}{ll} {\rm arbitrary}  & 0< t <
\epsilon \\ 2 \frac{t{\hat x}_1 (t) - \int_0^t\left[{\hat
 x}_1 (\sigma) - \sigma(\frac{c}{A}
\sqrt{{\hat x_1}(\sigma)}-\frac{1}{A}u(\sigma))\right]d\sigma}{t^2}
& \epsilon < t < t_I
\end{array} \right.
\]
where $\epsilon > 0$ is ``very small". The estimated value
$\hat{x}_1 (t)$ of $x_1 (t)$, which is obtained from Formula
(\ref{x1}), needs as in Section \ref{controller} the on-line
estimation $y_e (t)$ and ${\dot y}_e (t)$.

The estimated value $\hat{\bf w}$ of $\bf w$, which is detectable
and algebraically isolable (cf. Section \ref{pareq}), follows from
\begin{eqnarray}
\hat{\bf w}&=&1-\frac{1}{u (t)}\Big(2A\big(\frac{A}{c}{\dot y}_e (t)
+\sqrt{y_e (t)}\big) \nonumber \\ && \times \big(\frac{A}{c} {\ddot
y}_e (t) + \frac{{\dot y}_e (t)}{2\sqrt{y_e (t)}}\big)
\nonumber\\&&+c\big(\frac{A}{c}{\dot y}_e (t) +\sqrt{y_e
(t)}\big)-A{\hat\varpi}\Big) \nonumber
%
\end{eqnarray}

\subsection{Numerical simulations}
\begin{figure*}[htb]
\centering{{\rotatebox{-90}{\includegraphics[width=1.25\columnwidth]{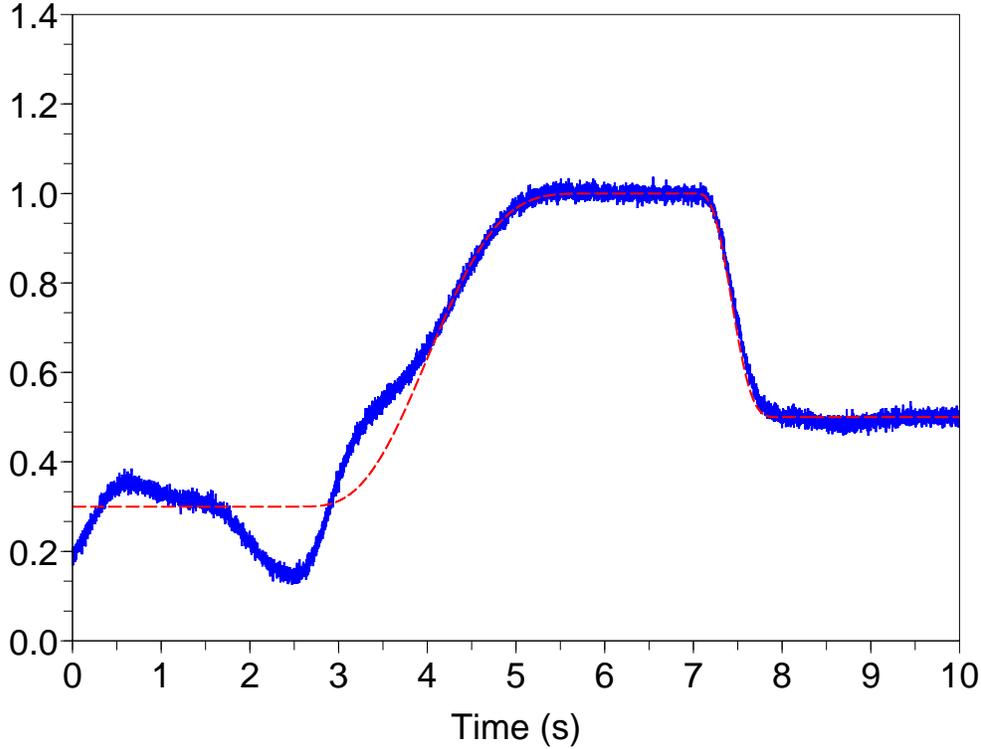}}}}
 \caption{$y^{\star}(t)$ (- -) and $y(t)$ (--) with fault accommodation\label{Nlinear2}}
\end{figure*}

\begin{figure}[htb]
\centering{{\rotatebox{-90}{\includegraphics[width=0.8\columnwidth]{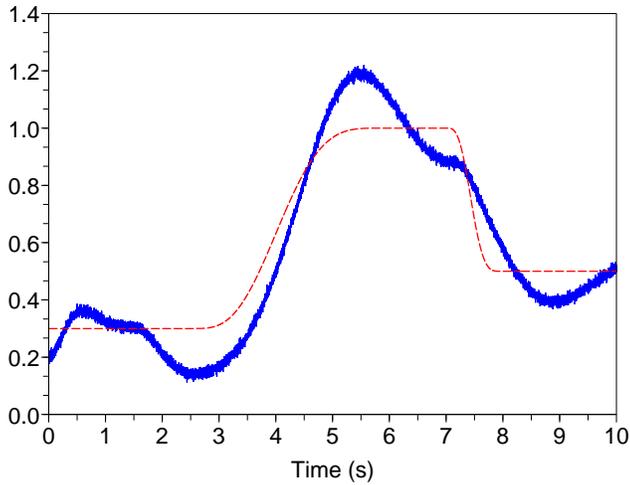}}}}
 \caption{$y^{\star}(t)$ (- -) and $y(t)$ (--) without fault accommodation\label{Nlinear1}}
\end{figure}

Figure \ref{Nlinear2} shows the closed-loop performance of our
trajectory tracking controller. The simulation scenario is the
following:
\begin{itemize}
\item The actuator fault ${\bf w}=0.7$
occurs at time $t_I = 1.5s$.
\item We estimate before the unknown constant perturbation $\varpi=0.2$ and use it
for estimating $\bf w$.
\item The fault tolerant control becomes effective at time $t=2.5s$.
\end{itemize}
Robustness is checked via an additive Gaussian white noise
$N(0;0.01)$. Comparison between Figures \ref{Nlinear1} and
\ref{Nlinear2} confirms the efficiency of our fault accommodation.

\section{Perturbation attenuation}\label{pertatt}

\subsubsection{Linear case\label{Ale}} Suppose we are given a
linear perturbed second order system
\begin{equation}
\ddot{y} (t) + y(t) = u(t) - z(t) + C \mathbf{1}(t - t_I)
\label{1exlin}
\end{equation}
where
\begin{itemize}
\item $z(t)$ is an unknown perturbation input,
\item $\mathbf{1}(t)$ is the Heaviside step function, i.e.,
$$
\mathbf{1}(t) = \left\{\begin{array}{l} 0 \quad \mbox{\rm if} \quad t < 0 \\
1 \quad \mbox{\rm if} \quad t \geq 0 \end{array} \right.
$$
\item $C$ is an unknown constant and thus $C \mathbf{1}(t - t_I)$
is a constant bias, of unknown amplitude, starting at time $t_I \geq
0$.
\end{itemize}
\begin{remark}
The difference $C \mathbf{1}(t - t_I) - z(t)$ is a rationally
determinable variable according to Section \ref{determin}.
\end{remark}
The estimate $z_e(t)$ of $z(t)$ is given up to a piecewise constant
error by
$$
z_e(t)=-\ddot{y}_e(t) - y_e(t)+u(t)
$$
where $y_e(t)$ and $\ddot{y}_e (t)$ are the on-line estimated values
of $y(t)$ and $\ddot{y} (t)$. We design a
generalized-proportional-integral (GPI) regulator, in order to track
asymptotically a given output reference trajectory $y^\star (t)$,
i.e.,
\begin{equation}
\label{gpi1} u(t) = y_e(t) + z_e(t) + {\ddot y}^{\star}(t)+
\mathcal{G} \star (y_e(t)-y^{\ast}(t))
\end{equation}
where \begin{itemize}
\item $\mathcal{G}$ is defined via its rational transfer function
$\frac{c_2s^2+c_1s+c_0}{s(s+c_3)}$
\item $s^4+ c_3s^3+ c_2s^2+ c_1s+ c_0$ is the
characteristic polynomial of the unperturbed closed-loop system. The
coefficients $c_0, c_1, c_2, c_3$ are chosen so that the imaginary
parts of its roots are strictly negative.
\end{itemize}
Like usual proportional-integral-derivative (PID) regulators, this
controller is robust with respect to un-modeled piecewise constant
errors

The computer simulations were performed with
\[
z(t)=\frac{10t^3\sin(2t)}{1+t^2+t^3}
\]
The unknown constant perturbation suddenly appears at time $t_I = 4$
with a permanent value $C = 1.25$. The coefficients of the
characteristic polynomial were forced to be those of the desired
polynomial $P_d(s)=(s^2+2\zeta\omega_ns+\omega_n^2)^2$, with $\zeta
= 0.81$, $\omega_n = 4$. We have set $y^{\star}(t)=\sin \omega t$,
$\omega =
2.5$[rad/s].\\
\begin{figure}[htb]
\centering{{\rotatebox{-90}{\includegraphics[width=0.81\columnwidth]{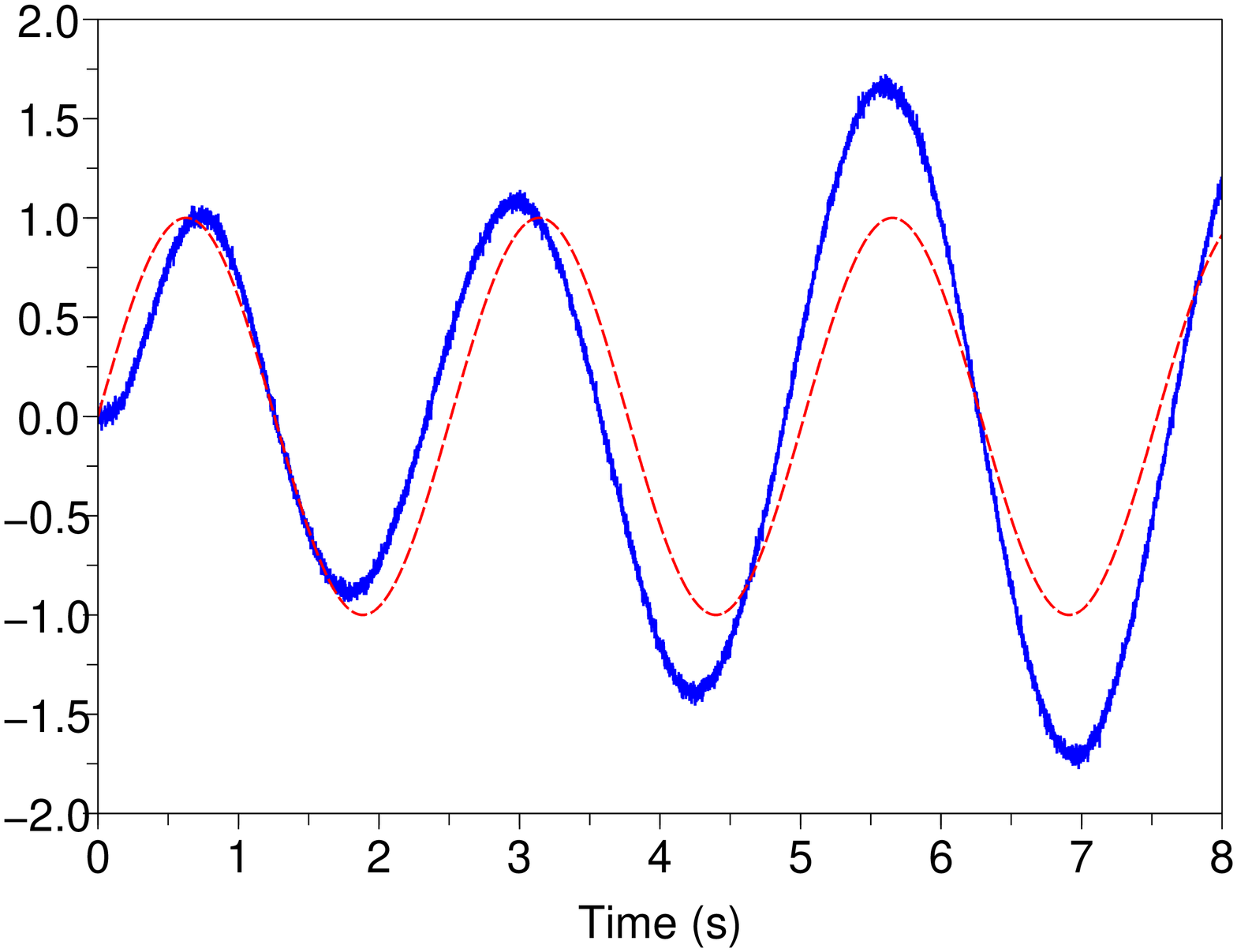}}}}
 \caption{$y^{\star}(t)$ (- -) and $y(t)$ (--) without perturbation attenuation \label{linear1}}
\end{figure}
\begin{figure*}[htb]
\centering{{\rotatebox{-90}{\includegraphics[width=1.25\columnwidth]{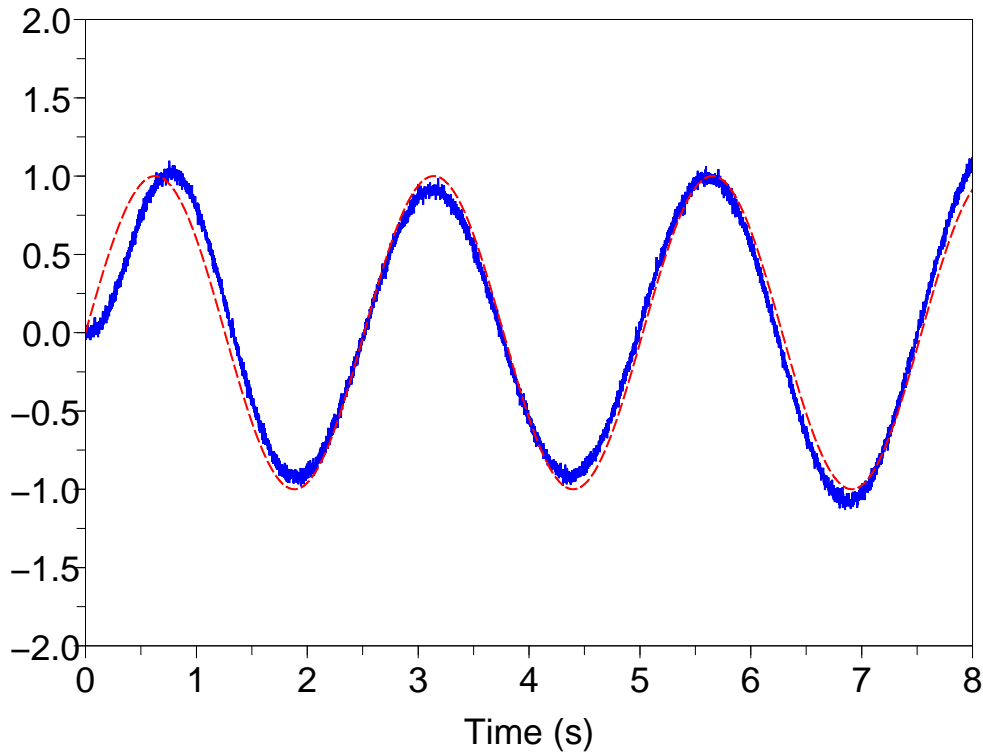}}}}
 \caption{$y^{\star}(t)$ (- -) and $y(t)$ (--) with perturbation attenuation \label{linear2}}
\end{figure*}
Figure \ref{linear1} (resp. \ref{linear2}) shows the reference
signal $y^{\star}(t)$ and the output signal $y(t)$ without
estimating $z_e (t)$ (resp. with the estimate $z_e (t)$). We added
in the simulations of Figure \ref{linear2} a Gaussian white noise
$N(0;0.025)$ to the measurement $y(t)$. The results are quite
remarkable.

\begin{remark}
The same technique yields an efficient solution to fault tolerant
linear control, which completes \cite{diag-lin}. Just think at
$z(t)$ as a fault variable.
\end{remark}

\subsubsection{Non-linear extension}

\begin{figure}[htb]
\centering{{\rotatebox{-90}{\includegraphics[width=0.81\columnwidth]{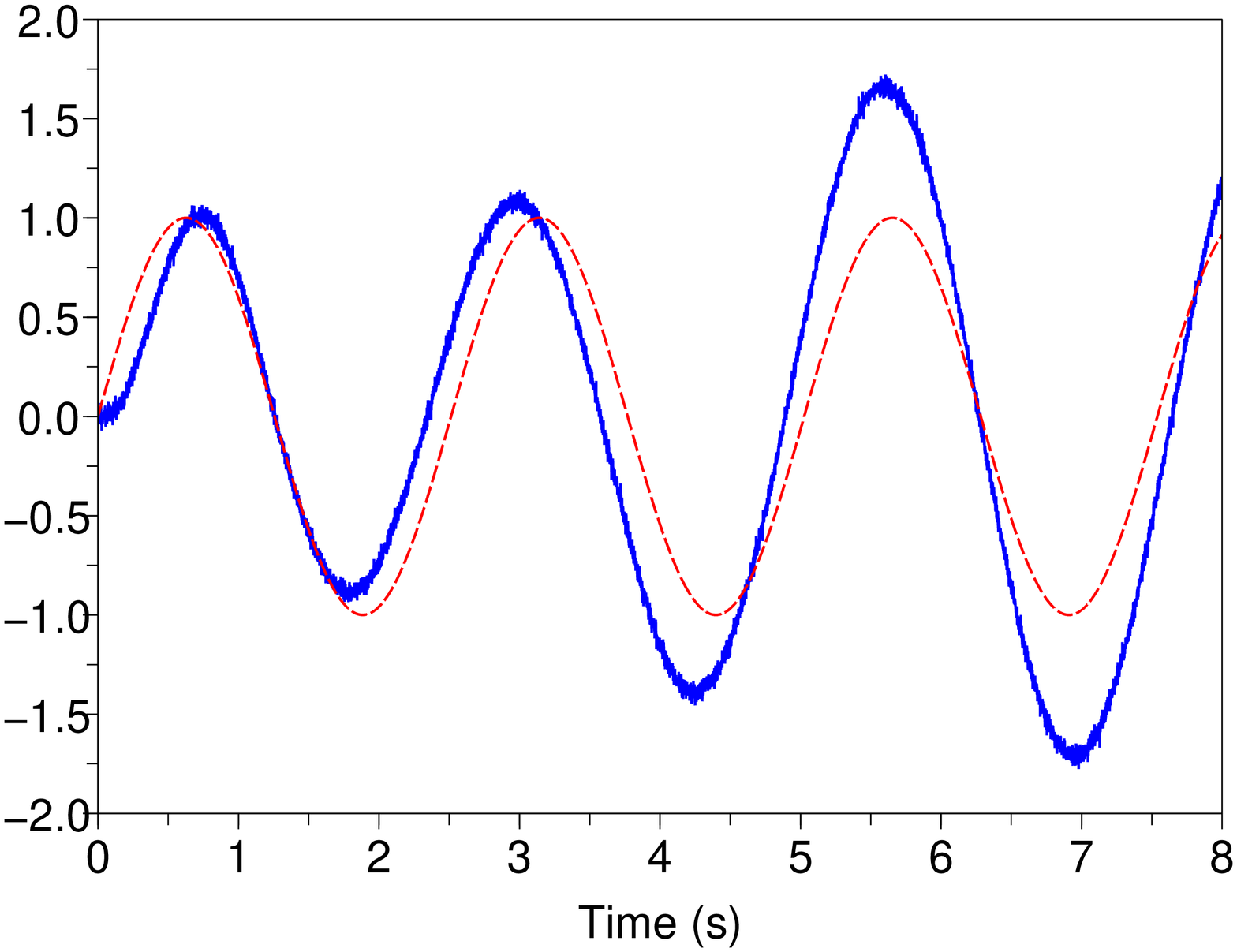}}}}
\caption{$y^{\star}(t)$ (- -) and $y(t)$ (--) without perturbation
attenuation \label{Nlinear3}}
\end{figure}

\begin{figure*}[htb]
\centering{{\rotatebox{-90}{\includegraphics[width=1.25\columnwidth]{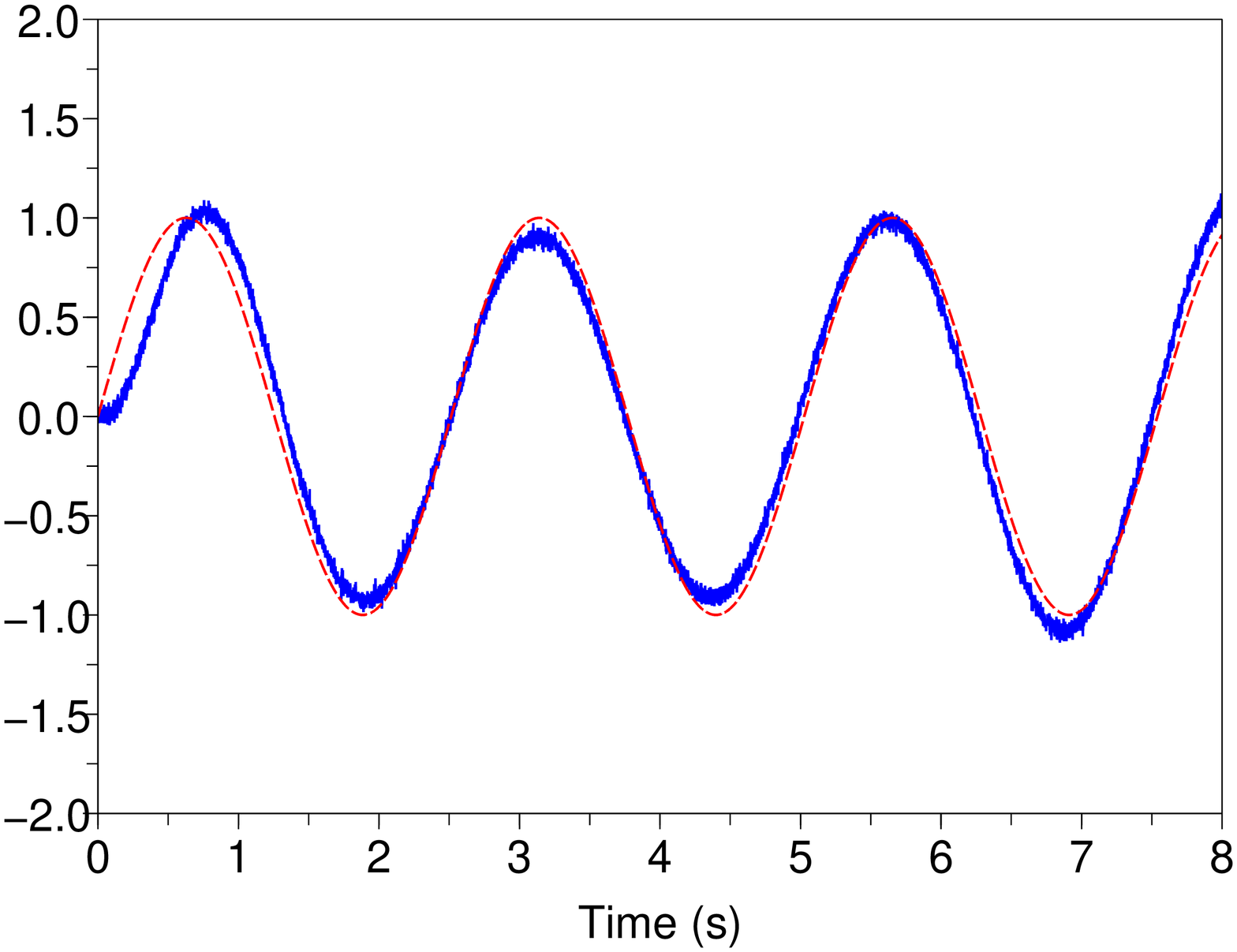}}}}
\caption{$y^{\star}(t)$ (- -) and $y(t)$ (--) with perturbation
attenuation.
 \label{Nlinear4}}
\end{figure*}

Replace the term $y(t)$ in system (\ref{1exlin}) by the product
$y(t) {\dot y}(t)$:
\begin{equation}\label{nl}
\ddot{y} (t) + y(t){\dot y}(t) = u(t) - z(t) + C \mathbf{1}(t - t_I)
\end{equation}
The perturbations $z(t)$ and $C \mathbf{1}(t - t_I)$ are the same as
above. The estimate $z_e (t)$ of $z(t)$ up to a piecewise constant
is given by
\[
z_e(t)=-{\ddot y}_e(t)-y_e{\dot y}_e(t)+u(t)
\]
where $y_e(t)$, ${\dot y}_e (t)$ and ${\ddot y}_e (t)$ are the
estimates of $y(t)$, ${\dot y} (t)$ and ${\ddot y} (t)$. The
feedback law (\ref{gpi1}) becomes
\begin{equation}\label{gpinl}
u=y_e(t)\dot y_e(t)+z_e(t) + {\ddot y}^{\star}(t)+ \mathcal{G} \star
(y_e(t)-y^{\ast}(t))
\end{equation}
\begin{remark}
Rewrite system (\ref{nl}) via the following state variable
representation
$$
\left\{ \begin{array}{l} {\dot x}_1 (t) =  x_2 (t)
\\ {\dot x}_2 (t) =  - x_1 (t) x_2 (t) + u(t) - z(t) + C \mathbf{1}(t - t_I)
\\ y (t) = x_1 (t)
\end{array} \right.
$$
Applying the feedback law (\ref{gpinl}) amounts possessing good
estimates of the two state variables.
\end{remark}
\noindent Figures \ref{Nlinear3} and \ref{Nlinear4} depict the
computer simulations with the same numerical conditions as before.
The results are again excellent.



\section{Conclusion}
We have proposed a new approach to non-linear estimation, which is
not of asymptotic nature and does not necessitate any statistical
knowledge of the corrupting noises\footnote{Let us refer to a recent
book by \cite{smolin}, which contains an exciting description of the
competition between various theories in today's physics. Similar
studies do not seem to exist in control.}. Promising results have
already been obtained, which will be supplemented in a near future
by other theoretical advances (see, e.g., \cite{barbot} on observers
with unknown inputs) and several concrete case-studies (see already
\cite{garcia,nothen}). Further numerical improvements will also be
investigated (see already \cite{mboup}).

\sectiona{ACKNOWLEDGMENT} Two authors (MF \& CJ) wish to
thank M. Mboup for a most fruitful cooperation on numerical
differentiation.

\bibliographystyle{apalike}

\end{document}